\documentclass[letterpaper,11pt]{article}
\usepackage{jheppub}

\usepackage{braket}
\usepackage{tikz}
\usepackage{hyperref}
\usepackage{amssymb}
\usepackage{amsmath}
\usepackage{color}

\DeclareMathOperator{\tr}{tr}

\newcommand{\nn}{\\ \nonumber} 
\newcommand \C {\mathcal{C}}

\newcommand \Open {\mathcal{O}}
\newcommand \mathtikz[1] {\quad \vcenter{\hbox{\tikz{#1}}} \quad}

\newcommand\idC[2] { 
\begin{scope}[xshift=#1,yshift=#2]
\filldraw[left color=lightgray, right color=white] (-0.25,0) -- (0.25,0) -- (0.25,-1) to [in=-90,out=-90] (-0.25,-1) -- (-0.25,0);
\filldraw[left color=white,right color=lightgray] (0,0) ellipse (0.25 and 0.1);, 
\draw[dotted] (0.25,-1) arc (0:180:0.25 and 0.1);
\end{scope}
}

\newcommand\idCshifted[3] { 
\begin{scope}[xshift=#1,yshift=#2]
\filldraw[left color=lightgray, right color=white] (#3-0.25,0) -- (#3+0.25,0) to [out=-90,in=90] (0.25,-1) to [in=-90,out=-90] (-0.25,-1) to [out=90,in=-90] (#3-0.25,0);
\filldraw[left color=white,right color=lightgray] (#3,0) ellipse (0.25 and 0.1);, 
\draw[dotted] (0.25,-1) arc (0:180:0.25 and 0.1);
\end{scope}
}

\newcommand\idA[2] { 
\begin{scope}[xshift=#1,yshift=#2]
\filldraw[fill=white,draw=black] (-0.25,0) rectangle (0.25,-1);
\end{scope}
}

\newcommand\muC[2]{ 
\begin{scope}[xshift=#1,yshift=#2]
\filldraw[left color=lightgray, right color=white] (-0.25,0) to [out=-90,in=180] (0,-0.33) to [in=-90,out=0] (0.25,0) to  (0.75,0) to [in=90,out=-90] (0.25,-1) to [out=-90,in=-90] (-0.25,-1) to [in=-90,out=90] (-0.75,0);
\filldraw[left color=white,right color=lightgray] (-0.5,0) ellipse (0.25 and 0.1);
\filldraw[left color=white,right color=lightgray] (0.5,0) ellipse (0.25 and 0.1);
\draw[dotted] (0.25,-1) arc (0:180:0.25 and 0.1);
\end{scope}
}

\newcommand\pairC[2]{ 
\begin{scope}[xshift=#1,yshift=#2]
\filldraw[left color=lightgray, right color=white] (-0.25,0) to [out=-90,in=180] (0,-0.33) to [in=-90,out=0] (0.25,0) to  (0.75,0) to [out=-90,in=0] (0, -0.83) to [out=180,in=-90]
(-0.75,0);
\filldraw[left color=white,right color=lightgray] (-0.5,0) ellipse (0.25 and 0.1);
\filldraw[left color=white,right color=lightgray] (0.5,0) ellipse (0.25 and 0.1);
\end{scope}
}

\newcommand\widecopairC[2]{ 
\begin{scope}[xshift=#1,yshift=#2]
\filldraw[left color=lightgray, right color=white] (-0.25,0) to [out=90,in=180] (0.25,0.33) to [in=90,out=0] (0.75,0) to [out=-90,in=-90] (1.25,0) to [out=90,in=0] (0.25, 0.83) to [out=180,in=90]
(-0.75,0) to [out=-90,in=-90] (-0.25,0);
\draw[dotted] (-0.25,0) arc (0:180:0.25 and 0.1);
\draw[dotted] (1.25,0) arc (0:180:0.25 and 0.1);
\end{scope}
}

\newcommand\copairC[2]{ 
\begin{scope}[xshift=#1,yshift=#2]
\filldraw[left color=lightgray, right color=white] (-0.25,0) to [out=90,in=180] (0,0.33) to [in=90,out=0] (0.25,0) to [out=-90,in=-90] (0.75,0) to [out=90,in=0] (0, 0.83) to [out=180,in=90]
(-0.75,0) to [out=-90,in=-90] (-0.25,0);
\draw[dotted] (-0.25,0) arc (0:180:0.25 and 0.1);
\draw[dotted] (0.75,0) arc (0:180:0.25 and 0.1);
\end{scope}
}

\newcommand\deltaC[2]{
\begin{scope}[xshift=#1,yshift=#2]
\filldraw[left color=lightgray, right color=white] (-0.25,-1) to [out=90,in=180] (0,-0.66) to [in=90,out=0] (0.25,-1) to [out=-90,in=-90] (0.75,-1) to [in=-90,out=90] (0.25,0) to (-0.25,0) to [in=90,out=-90] (-0.75,-1) to [out=-90,in=-90] (-0.25,-1);
\filldraw[left color=white,right color=lightgray] (0,0) ellipse (0.25 and 0.1);
\draw[dotted] (-0.25,-1) arc (0:180:0.25 and 0.1);
\draw[dotted] (0.75,-1) arc (0:180:0.25 and 0.1);
\end{scope}
}

\newcommand\muA[2]{ 
\begin{scope}[xshift=#1,yshift=#2]
\draw (-0.75,0) -- (-0.25,0) to [out=-90,in=180] (0,-0.33) to [in=-90,out=0] (0.25,0) -- (0.75,0) to [in=90,out=-90] (0.25,-1);
\draw (-0.25,-1) -- (0.25,-1);
\draw (-0.75,0) to [in=90,out=-90] (-0.25,-1);
\end{scope}
}

\newcommand\pairA[2]{ 
\begin{scope}[xshift=#1,yshift=#2]
\draw (-0.75,0) -- (-0.25,0) to [out=-90,in=180] (0,-0.33) to [in=-90,out=0] (0.25,0) -- (0.75,0) to [out=-90,in=0] (0,-0.83) to [out=180,in=-90] (-0.75,0);
\end{scope}
}

\newcommand\copairA[2]{ 
\begin{scope}[xshift=#1,yshift=#2]
\draw (-0.75,0) -- (-0.25,0) to [out=90,in=180] (0,0.33) to [in=90,out=0] (0.25,0) -- (0.75,0) to [out=90,in=0] (0,0.83) to [out=180,in=90] (-0.75,0);
\end{scope}
}

\newcommand\deltaA[2]{ 
\begin{scope}[xshift=#1,yshift=#2]
\draw (-0.75,-1) -- (-0.25,-1) to [out=90,in=180] (0,-0.66) to [in=90,out=0] (0.25,-1) -- (0.75,-1) to [in=-90,out=90] (0.25,0) -- (-0.25,0) to [in=90,out=-90] (-0.75,-1);
\end{scope}
}

\newcommand\widedeltaA[2]{ 
\begin{scope}[xshift=#1,yshift=#2]
\draw (-1.25,-1) -- (-0.75,-1) to [out=90,in=180] (0,-0.66) to [in=90,out=0]
(0.75,-1) -- (1.25,-1) to [in=-90,out=90] (0.25,0) -- (-0.25,0) to [in=90,out=-90] (-1.25,-1);
\end{scope}
}

\newcommand\semiwidedeltaA[2]{ 
\begin{scope}[xshift=#1,yshift=#2]
\draw (-1,-1) -- (-0.5,-1) to [out=90,in=180] (0,-0.66) to [in=90,out=0]
(0.5,-1) -- (1,-1) to [in=-90,out=90] (0.25,0) -- (-0.25,0) to [in=90,out=-90] (-1,-1);
\end{scope}
}

\newcommand\semiwidemuA[2]{ 
\begin{scope}[xshift=#1,yshift=#2,yscale=-1]
\draw (-1,-1) -- (-0.5,-1) to [out=90,in=180] (0,-0.66) to [in=90,out=0]
(0.5,-1) -- (1,-1) to [in=-90,out=90] (0.25,0) -- (-0.25,0) to [in=90,out=-90] (-1,-1);
\end{scope}
}

\newcommand\zipper[2]{ 
\begin{scope}[xshift=#1,yshift=#2]
\draw (-0.25,-1) -- (0.25,-1);
\filldraw[right color=white,left color=lightgray] (-0.25,0) to (-0.25,-1) to [out=90,in=225] (0,-0.5) to [out=-45,in=90] (0.25,-1) to (0.25,0);
\filldraw[left color=white,right color=lightgray] (0,0) ellipse (0.25 and 0.1);
\end{scope}
}

\newcommand\cozipper[2]{ 
\begin{scope}[xshift=#1,yshift=#2]
\draw (-0.25,0) -- (0.25,-0);
\filldraw[right color=white,left color=lightgray] (-0.25,-1) to (-0.25,0) to [out=-90,in=135] (0,-0.5) to [out=45,in=-90] (0.25,0) to (0.25,-1) to [in=-90,out=-90] (-0.25,-1);
\draw[dotted] (0.25,-1) arc (0:180:0.25 and 0.1);
\end{scope}
}

\newcommand\epsilonC[2]{ 
\begin{scope}[xshift=#1,yshift=#2]
\filldraw[right color=white,left color=lightgray] (-0.25,0) to [out=-90,in=180] (0,-0.33) to [in=-90,out=0] (0.25,0);
\filldraw[left color=white,right color=lightgray] (0,0) ellipse (0.25 and 0.1);
\end{scope}
}

\newcommand\etaC[2] { 
\begin{scope}[xshift=#1,yshift=#2]
\filldraw[right color=white,left color=lightgray] (-0.25,0) to [out=90,in=180] (0,0.33) to [in=90,out=0] (0.25,0) to [in=-90,out=-90] (-0.25,0);
\draw[dotted] (0.25,0) arc (0:180:0.25 and 0.1);
\end{scope}
}

\newcommand\epsilonA[2] {
\begin{scope}[xshift=#1,yshift=#2]
\draw (-0.25,0) -- (0.25,0);
\draw (-0.25,0) to [out=-90,in=180] (0,-0.33) to [in=-90,out=0] (0.25,0);
\end{scope}
}

\newcommand\etaA[2] {
\begin{scope}[xshift=#1,yshift=#2]
\draw (-0.25,0) -- (0.25,0);
\draw (-0.25,0) to [out=90,in=180] (0,0.33) to [in=90,out=0] (0.25,0);
\end{scope}
}

\newcommand\tauA[2] { 
\begin{scope}[xshift=#1,yshift=#2]
\draw (-0.75,0) -- (-0.25,0) to [out=-90,in=90,looseness=0.5] (0.75,-1) -- (0.25,-1) to [out=90,in=-90,looseness=0.5] (-0.75,0);
\filldraw[fill=white,draw=black] (0.75,0) -- (0.25,0) to [out=-90,in=90,looseness=0.5] (-0.75,-1) -- (-0.25,-1) to [out=90,in=-90,looseness=0.5] (0.75,0);
\end{scope}
}

\newcommand\tauC[2] {
\begin{scope}[xshift=#1,yshift=#2]
\filldraw[right color=white,left color=lightgray] (-0.75,0) -- (-0.25,0) to  [out=-90,in=90,looseness=0.5] (0.75,-1) to [in=-90,out=-90] (0.25,-1) to [out=90,in=-90,looseness=0.5] (-0.75,0);
\filldraw[right color=white,left color=lightgray] (0.75,0) -- (0.25,0) to [out=-90,in=90,looseness=0.5] (-0.75,-1) to [in=-90,out=-90] (-0.25,-1) to [out=90,in=-90,looseness=0.5] (0.75,0);
\filldraw[left color=white,right color=lightgray] (-0.5,0) ellipse (0.25 and 0.1);
\filldraw[left color=white,right color=lightgray] (0.5,0) ellipse (0.25 and 0.1);
\draw[dotted] (-0.25,-1) arc (0:180:0.25 and 0.1);
\draw[dotted] (0.75,-1) arc (0:180:0.25 and 0.1);
\end{scope}
}

\begin{document}

\title{Entanglement branes, modular flow, and extended topological quantum field theory}

\author[a]{William Donnelly}
\emailAdd{wdonnelly@perimeterinstitute.ca}
\author[b]{Gabriel Wong}
\emailAdd{gabrielwon@gmail.com}
\abstract{
Entanglement entropy is an important quantity in field theory, but its definition poses some challenges.
The naive definition involves an extension of quantum field theory in which one assigns Hilbert spaces to spatial sub-regions.
For two-dimensional topological quantum field theory we show that the appropriate extension is the open-closed topological quantum field theory of Moore and Segal.
With the addition of one additional axiom characterizing the ``entanglement brane'' we show how entanglement calculations can be cast in this framework.
We use this formalism to calculate modular Hamiltonians, entanglement entropy and negativity in two-dimensional Yang-Mills theory and relate these to singularities in the modular flow.
As a byproduct we find that the negativity distinguishes between the ``log dim $R$'' edge term and the ``Shannon'' edge term.
We comment on the possible application to understanding the Bekenstein-Hawking entropy in two-dimensional gravity.
}

\maketitle

\section{Introduction}

Entanglement is an important quantity in field theory, but its precise definition carries subtleties.
Naively, one associates to a division of space into two parts $A$ and $B$ with a factorization of Hilbert spaces
\begin{equation}
    \mathcal{H}_{A \cup B} = \mathcal{H}_A \otimes \mathcal{H}_B.
\end{equation}
This factorization holds for scalar fields with a lattice regulator, but one has to be more careful in the continuum.
The basic issue is that while quantum field theory naturally comes equipped with a Hilbert space associated with a Cauchy surface, it does not naturally associate Hilbert spaces $\mathcal{H}_A$ to regions with boundary.

One issue is that in the algebraic approach to quantum field theory, regions of space are associated with von Neumann algebras, rather than Hilbert space factors.
This is essentially an ultraviolet issue: quantum field theory only allows for nonsingular states, which constrains the form of the two-point function at short distances.
While the entanglement entropy is not well-defined in this context, it is still possible to define quantities such as relative entropy in the algebraic setting \cite{Araki:1976zv} that are ultraviolet finite \cite{Casini:2008cr}. 
See Ref.~\cite{Witten:2018lha} for a recent review of the algebraic approach to entanglement. 

While the above issues pertain to continuum quantum field theory, much of the interest in entanglement entropy comes from its relation to the Bekenstein-Hawking entropy and therefore to quantum gravity.
At short distances we expect to exit the regime of validity of quantum field theory on a fixed background.
In the quantum gravity description it is unclear whether local algebras of observables exist, or what their classification as von Neumann algebras would be.
This is a consequence of diffeomorphism invariance, and is essentially an infrared issue.

In gauge theory, the physical Hilbert space consists of wavefunctionals satisfying local constraints.
As a result, the Hilbert space does not have a local tensor product structure \cite{Buividovich2008b,Donnelly:2011hn, Casini:2013rba,Donnelly:2014gva,Ghosh:2015iwa,Hung:2015fla,Soni:2015yga,Lin:2018bud,Blommaert:2018rsf,Blommaert:2018oue}.
Instead, one can associate an extended Hilbert space to each region of space which contains edge modes on the boundary.
The edge modes carry gauge charges which allow Wilson lines to end on the boundary.
These local Hilbert spaces can be combined with an entangling product \cite{Donnelly:2016auv}, which enforces cancellation of the surface charges.

The purpose of the present article is to show how this generalized notion of entanglement fits naturally within the context of topological quantum field theory (TQFT).
In the axiomatic formulation of closed TQFT one associates Hilbert spaces with closed, codimension-1 manifolds and disjoint unions with their tensor products \cite{Atiyah:1989vu}. 
The evolution of these manifolds is described by cobordisms, which are assigned to linear maps. 
These assignments arise from computing the Euclidean path integral on a cobordism, and by gluing a basic set of cobordisms one can obtain the path integral on a general manifold.

To describe entanglement of regions within a single connected spacetime, we need additional rules for describing the Hilbert space of manifolds with codimension-2 boundaries, which we identify as entangling surfaces.  
This leads to a richer set of cobordisms arising from cutting manifold along codimension-1 as well as codimension-2 surfaces. 
An \emph{extended} TQFT is the mathematical framework that describes cutting and gluing of manifolds along surfaces of arbitrary codimension.  
In particular in two-dimensions, Moore and Segal derived sewing axioms that ensure the compatibility of different ways of cutting the same manifold.
The sewing axioms were meant to classify D-branes, viewed as objects in the category of boundary conditions; in this work we interpret them as rules that classify extended Hilbert spaces and their edge modes. 

To describe entanglement in the extended TQFT formalism, we have to formulate the extended Hilbert space construction as a \emph{spacetime} process.  
In particular, the rule for embedding the Hilbert space of a circle into that of an interval, and for embedding the Hilbert space of one interval into a larger interval are described by cobordisms:
\begin{equation} \label{splitting}
\mathtikz{ \zipper{0cm}{0cm} }: \mathcal{H}_\text{circle} \to \mathcal{H}_\text{interval}, \qquad
\mathtikz{ \deltaA{0cm}{0cm} }:
\mathcal{H}_\text{interval} \to \mathcal{H}_\text{interval} \otimes \mathcal{H}_\text{interval}. \qquad
\end{equation}
These diagrams are to be read from top to bottom, and describe a circle being cut open into an interval, and that interval being split into two subintervals.
By repeating these maps we can view a state of the circle as a state in the tensor product of any number of intervals.

Each time we apply one of the splitting rules \eqref{splitting} we introduce a new codimension-1 boundary around the entangling surface.
To ensure that the introduction of the entangling surface does not change the state, we require that holes in the diagrams can be sewn up:
\begin{equation}\label{holes}
\mathtikz{ \cozipper{0cm}{0cm} \zipper{0cm}{1cm} } 
= \mathtikz{ \idC{0cm}{0cm} },
\qquad
\mathtikz{ \muA{0cm}{0cm} \deltaA{0cm}{1cm} } = \mathtikz{ \idA{0cm}{0cm} } .
\end{equation}
This is a boundary condition that was identified in \cite{Donnelly:2016jet} as an entanglement brane.  In a different context, this boundary condition has been used to obtain integrable lattice models from line operators in TQFT's (see \cite{ Yagi:2016oum} and the references within) 

We begin in section \ref{section:tqft} by reviewing the axioms of ``open-closed'' TQFT and its diagrammatic notation.
We will avoid discussion of the underlying category theory, details of which can be found elsewhere \cite{Lauda:2005wn}.
In section \ref{section:ebrane} we show how entanglement can be described in open-closed TQFT upon introducing the entanglement brane axiom, which allows us to sew up holes as in \eqref{holes}.
We show how this can be used to study entanglement entropy, modular flows and negativity of states produced by Euclidean path integrals on arbitrary Riemann surfaces.

In sections \ref{section:ym} and \ref{section:string} we consider the specific example of two-dimensional Yang-Mills theory, and the closely related chiral Gross-Taylor string theory.
While not strictly topological, these theories can be treated using TQFT methods.
We show how each of these theories can be cast as open-closed TQFTs satisfying the entanglement brane axiom.  
In the case of two-dimensional Yang-Mills, we apply this formalism compute entanglement entropy, modular Hamiltonians, as well as negativity of general subregions and states, generalizing results of \cite{Donnelly:2014gva}. 
For the chiral Gross-Taylor string, we provide a worldsheet interpretation of the entanglement brane in the case of arbitrary entangling surfaces and for general states. 
This provides a worldsheet prescription for calculations of entanglement entropy and related quantities.

\section{Open-Closed TQFT}
\label{section:tqft}
The diagrammatic structure of an open-closed TQFT originated from the factorization of string worldsheet amplitudes that describe interactions of open and closed strings \cite{Lewellen:1991tb}.   Here we will review the subject as formulated by 
\cite{MooreTalk,Lauda:2005wn,Moore:2006dw}. 
A nice informal treatment is given by \cite{Baez}.

\subsection{Closed TQFT}
 A two dimensional closed TQFT is a rule that assigns a vector space $\otimes^{n} \mathcal{C}$ to the disjoint union of $n$ circle and linear maps to cobordisms between circles.  The circles are oriented, and a change of orientation corresponds to taking the dual of the vector space. Gluing cobordisms then corresponds to composition of linear maps. The ``pair of pants" cobordism given by $\mu_{\mathcal{C}}$ in \eqref{closed} defines a multiplication on $\mathcal{C}$ that endows it with the structure of an algebra. The equivalence of cobordisms related by orientation preserving diffeomorphisms imply relations that make a two dimensional closed TQFT a commutative Frobenius algebra.

A Frobenius algebra is an algebra $\C$ with some additional operations:
\begin{align}
    \mu &: \C \otimes \C \to \C, & \text{product} \\
    \eta &: \mathbb{C} \to \C, & \text{unit} \\
    \Delta &: \C \to \C \otimes \C, & \text{coproduct} \\
    \epsilon &: \C \to \mathbb{C}, & \text{counit/trace}
\end{align}
There is also a braiding operation $\tau$, which just maps $X \otimes Y \to Y \otimes X$.
Using these one can construct a natural pairing $\pi = \epsilon \circ \mu: \C \otimes \C \to \mathbb{C}$. For a Frobenius algebra $\C$, this is non-degenerate and satisfies the invariance condition  
\begin{align} \label{inv}
    \pi(a b,c)=\pi(a,b c)
\end{align}

The algebraic operations correspond to a set of elementary cobordisms: 
\begin{align} \label{closed}
    \mu_\C = \!\! \mathtikz { \muC{0cm}{0cm} } \quad
    \eta_\C = \!\! \mathtikz{ \etaC{0cm}{0cm} } \quad
    \Delta_\C = \!\! \mathtikz { \deltaC{0cm}{0cm} } \quad
    \epsilon_\C = \!\! \mathtikz { \epsilonC{0cm}{0cm} } \quad
    \tau_\C = \!\! \mathtikz { \tauC{0cm}{0cm} }
\end{align}
In addition, it is useful to include the identity map:
\begin{align}
    \mathbf{1}= \mathtikz{\idC{0}{0}}
\end{align}
Given these definitions, the topological invariance of the TQFT ensures that it satisfies the rules of a Frobenius algebra. 
For example, the condition that $\eta_{\C}$ is the unit follows from 
\begin{align}
    \mathtikz{ \muC{.5cm}{1cm}
    \etaC{0cm}{ 1.cm}} = \mathtikz{\idC{0cm}{0cm}}
\end{align}

An arbitrary compact,oriented 2D manifold can be obtained by gluing the elementary cobordisms in \eqref{closed}. The compatibility of different gluings is ensured because of the
associativity
\begin{equation}
    \mathtikz{ \muC{0.5cm}{-1cm} \idCshifted{1cm}{0cm}{0.5} \muC{0cm}{0cm} } = \mathtikz{ \muC{0.5cm}{-1cm} \idCshifted{0cm}{0cm}{-0.5} \muC{1cm}{0cm} }
\end{equation}
and commutativity 
($\mu = \mu \circ \tau$):
\begin{equation}
\mathtikz{ \muC{0cm}{0cm} } = \mathtikz { \muC{0cm}{-0.5cm}  \tauC{0cm}{0.5cm} }
\end{equation}

The cobordism describing the pairing $\pi$ is:
\begin{equation} \label{pairing}
    \mathtikz{ \pairC{0cm}{0cm} } := \mathtikz{ \epsilonC{0cm}{0cm} \muC{0cm}{1cm} }
\end{equation}

The Frobenius condition
\begin{equation}
\mathtikz{ \deltaC{0}{-0.5cm} \muC{0}{0.5cm} }
= \mathtikz{ \idC{4.5cm}{-0.5cm} \muC{3cm}{-0.5cm} \deltaC{4cm}{0.5cm} \idC{2.5cm}{0.5cm} }
= \mathtikz{ \idC{6.5cm}{-0.5cm}\muC{8cm}{-0.5cm}\deltaC{7cm}{0.5cm}\idC{8.5cm}{0.5cm} }
\end{equation}
then implies the zigzag identity, obtained by
attaching a unit to the above diagrams:
\begin{equation}
\mathtikz{ \pairC{0cm}{0cm}; \idC{1.5cm}{0cm}; \copairC{1cm}{0cm}; \idC{-0.5cm}{1cm}; } = \mathtikz{ \idC{0cm}{0cm} }.
\end{equation}
This expresses the fact that the pairing $\pi$ is nondegenerate.
The invariance condition \eqref{inv} follows from gluing a co-unit to the associativty constraint.  

\subsection{Open TQFT}
An open TQFT is similar to a closed TQFT except that the cobordisms are now oriented manifolds with boundaries.
Here the Hilbert spaces are associated to intervals, and the basic building blocks correspond to the diagrams:
\begin{align} \label{open}
    \mu_\Open = \!\! \mathtikz{ \muA{0cm}{0cm} } \quad
    \eta_\Open = \!\!\mathtikz{ \etaA{0cm}{0cm} } \quad
    \Delta_\Open = \!\!\mathtikz{ \deltaA{0cm}{0cm} } \quad
    \epsilon_\Open = \!\! \mathtikz{ \epsilonA{0cm}{0cm} } \quad
    \tau_\Open = \!\!\mathtikz{ \tauA{0cm}{0cm} }.
\end{align}
While the multiplication $\mu_{\mathcal{O}}$ is associative, it is \emph{not} commutative:
\begin{equation}
\mathtikz{ \muA{0cm}{0cm} } \neq 
\mathtikz { \muA{0cm}{-0.5cm}  \tauA{0cm}{0.5cm} }
\end{equation}
Instead we require the weaker property that it be symmetric ($\epsilon \circ \mu = \epsilon \circ \mu \circ \tau$):
\begin{equation}
\mathtikz { \epsilonA{0cm}{-1cm} \muA{0cm}{0cm} } = 
\mathtikz{ \epsilonA{0cm}{-1.5cm} \muA{0cm}{-0.5cm}  \tauA{0cm}{0.5cm} }
\end{equation}
which says that the bilinear form is symmetric,
\begin{equation} \label{symmetry}
    \mathtikz{ \pairA{0cm}{0cm} } = \mathtikz{ \pairA{0cm}{-1cm} \tauA{0cm}{0cm} }
\end{equation}
The Frobenius condition
\begin{equation}
\mathtikz{ \deltaA{0}{-0.5cm} \muA{0}{0.5cm} }
= \mathtikz{ \idA{4.5cm}{-0.5cm} \muA{3cm}{-0.5cm} \deltaA{4cm}{0.5cm} \idA{2.5cm}{0.5cm} }
= \mathtikz{ \idA{6.5cm}{-0.5cm}\muA{8cm}{-0.5cm}\deltaA{7cm}{0.5cm}\idA{8.5cm}{0.5cm} }
\end{equation}
holds, so an open TQFT is a \emph{symmetric} Frobenius algebra.

\subsection{Open-closed TQFT}

To describe a state on the circle in terms of states on an interval, we need a unified framework that includes both closed and open cobordisms.
This structure is known as an open-closed TQFT. 

Open-closed TQFTs are classified by \emph{knowledgeable Frobenius algebras}.
A knowledgeable Frobenius algebra is a combination of a commutative Frobenius algebra $\C$ (representing the closed sector) and a symmetric Frobenius algebra $\Open$ (representing the open sector). 
It also has two additional morphisms: the zipper $i: \C \to \Open$ and a dual cozipper $i^*: \Open \to \C$.
\begin{align}
    i &: \C \to \Open & \text{zipper} \\
    i^* &: \Open \to \C & \text{cozipper}
\end{align}
These are expressed graphically by the diagrams
\begin{equation} \label{zipper}
    i = \mathtikz { \zipper{0cm}{0cm} }, 
    \qquad 
    i^* = \mathtikz { \cozipper{0cm}{0cm} }
\end{equation}

There are some further consistency conditions that relate the open and closed sectors.

\begin{enumerate}
\item The zipper preserves the unit:
\begin{equation} \label{axiom1}
\mathtikz{ \zipper{0cm}{1cm} \etaC{0cm}{1cm} } = 
\mathtikz{ \etaA{0cm}{0cm} }
\end{equation}
\item The zipper preserves the product:
\begin{equation}
\tikz[baseline=0cm] { \zipper{0cm}{0cm} \muC{0cm}{1cm} } \quad = \quad 
\tikz[baseline=0cm] { \muA{0cm}{0cm} \zipper{-0.5cm}{1cm} \zipper{0.5cm}{1cm} }
\end{equation}
\item \textbf{Knowledge} The zipper maps into the center of the open string category, so the open strings "know" about the center.
\begin{equation}
\tikz[baseline=0cm] {\muA{0cm}{0cm} \zipper{-0.5cm}{1cm} \idA{0.5cm}{1cm} }
\quad = \quad
\tikz[baseline=0cm] { \muA{3cm}{-0.5cm} \tauA{3cm}{0.5cm} \zipper{2.5cm}{1.5cm} \idA{3.5cm}{1.5cm} }
\end{equation}
\item \textbf{Duality} The cozipper is dual to the zipper.
\begin{equation}
\mathtikz {
\pairA{0cm}{0cm}
\zipper{-0.5cm}{1cm} \idA{0.5cm}{1cm} }
\quad = \quad
\mathtikz {
\pairC{3cm}{0cm}
\idC{2.5cm}{1cm} \cozipper{3.5cm}{1cm} }
\end{equation}
\item \textbf{Cardy} The ``double twist'' projects onto the center.
\begin{equation}  \label{axiom5}
\mathtikz{ \muA{0}{-0.5cm} \tauA{0}{0.5cm} \deltaA{0}{1.5cm} }
=
\mathtikz{ \zipper{0cm}{0cm} \cozipper{0cm}{1cm} }
\end{equation}
\end{enumerate}

Note that the Cardy condition can be put into a more familiar form by sandwiching it between the open unit and counit and using the symmetry property \eqref{symmetry}. The result is:
\begin{equation}
\mathtikz{ \pairA{0cm}{0cm} \copairA{0cm}{0cm} } =
\mathtikz{ \epsilonA{0cm}{-2cm} \zipper{0cm}{-1cm} \cozipper{0cm}{0cm} \etaA{0cm}{0cm} }.
\end{equation}
On the left is the open string slicing in which the cylinder partition function is viewed as a trace, and on the right is the closed string slicing it is viewed as an amplitude between boundary states.

These conditions ensure topological invariance of the partition function: any manifold with boundary can be decomposed into the basic building blocks \eqref{closed}, \eqref{open} and \eqref{zipper} and the identities for the open and closed sectors, together with equations \eqref{axiom1} - \eqref{axiom5}.
This was proved in \cite{Lauda:2005wn}.

\subsection{Branes}
\label{section:branes} 
Given a closed TQFT, there can in general be multiple ways to extend it to an open/closed TQFT.
In the string theory description, this corresponds to the fact that there can be different types of branes on which the open strings can end.
In this case we can associate labels $a,b,c,\ldots$ to the boundaries of the open diagrams, with the rule that we can only compose morphisms when their boundary labels match.

For example, for each triple of labels $a,b,c$ we have a $\mu_{\Open,a,b,c}: \mathcal{H}_{ab} \otimes \mathcal{H}_{bc} \to \mathcal{H}_{ac}$ which we denote:
\begin{equation}
\mathtikz{ \muA{0cm}{0cm} 
\draw (-0.8cm,-0.5cm) node {\footnotesize $a$};
\draw (0cm,-0.1cm) node {\footnotesize $b$};
\draw (0.8cm,-0.5cm) node {\footnotesize $c$};
}
\end{equation}
The rule for composing such diagrams is that the labels have to match whenever they are joined via the boundary of an open string.

Thus when splitting a Hilbert space using the zipper or comultiplication, we have to make a choice of brane to insert at the entangling surface.
This is the subject of the next section.

\section{The entanglement brane}
\label{section:ebrane}

We now consider how to describe entanglement between regions of space in the formalism of open-closed TQFT.
As a simplest example we will consider the \emph{Hartle-Hawking state} $\ket{HH}$ which is the state produced by the unit cobordism:
\begin{equation} \label{hhstate}
    \ket{HH} = \mathtikz{ \etaC{0cm}{0cm} }.
\end{equation}
We would like to express this state as an entangled state of two intervals.
We can do this using the zipper \eqref{zipper} and the open coproduct \eqref{open}:
\begin{equation} \label{splitsphere}
\mathtikz{ \etaC{0cm}{0cm}} \to 
\mathtikz{ \zipper{0cm}{0cm} \etaC{0cm}{0cm} } \to 
\mathtikz{ \deltaA{0cm}{0cm} \zipper{0cm}{1cm} \etaC{0cm}{1cm} }
\end{equation}
This maps a state on the circle to an entangled state of two intervals.
By continuing to apply the open coproduct, we can decompose the state into an arbitrary number of intervals.

In general when doing the procedure \eqref{splitsphere} we have to make a choice of boundary condition at each step, so the state should really be denoted:
\begin{equation} \label{splitspherelabels}
\mathtikz{ 
\deltaA{0cm}{0cm} \zipper{0cm}{1cm} \etaC{0cm}{1cm}
\draw (-0.8cm,-0.5cm) node {\footnotesize $a$};
\draw (0cm,-0.9cm) node {\footnotesize $b$};
\draw (0.8cm,-0.5cm) node {\footnotesize $a$};
}\end{equation}
This state is different from the one we started with; the original state has no boundaries while the new one does.
The state therefore depends on the choice of boundary conditions, or more generally on the state inserted at the boundary.

The definition of the reduced density matrix of a subregion is that expectation values of operators restricted to that region calculated with the reduced density matrix agree with expectation values calculated in the original state.
For a partition of a system into parts $A$ and $B$, with a local operator $\mathcal{O}_A$ on system $A$ this means
\begin{equation} \label{partialtrace}
\tr_A[ \tr_B(\rho) \mathcal{O}_A ] = \tr[\rho (\mathcal{O}_A \otimes 1)]
\end{equation}
This constraint is actually quite powerful, and was used in \cite{Donnelly:2014gva} to argue for the presence of edge modes in the entanglement entropy of two-dimensional Yang-Mills theory.
Here we will see that it fixes the boundary condition associated to the entangling surface.

In order for property \eqref{partialtrace} to hold, expectation values in the state \eqref{splitspherelabels} should be the same as those in the original Hartle-Hawking state.
This is a condition on the boundary labels: we demand the existence of a label $e$ (for entanglement) such that
\begin{equation} \label{shrinkability}
\mathtikz{ \cozipper{0cm}{0cm} \zipper{0cm}{1cm} 
\draw (0cm,0.25cm) node {\footnotesize $e$};
\draw (0cm,-0.25cm) node {\footnotesize $e$};
} 
= \mathtikz{ \idC{0cm}{0cm} },
\qquad
\mathtikz{ \muA{0cm}{0cm} \deltaA{0cm}{1cm} 
\draw (0cm,0.15cm) node {\footnotesize $e$};
\draw (0cm,-0.15cm) node {\footnotesize $e$};
} = \mathtikz{ \idA{0cm}{0cm} } .
\end{equation}
Since the different labels of the boundaries correspond to branes, we call this boundary condition the entanglement brane, following \cite{Donnelly:2016jet}.\footnote{We will see in section \ref{section:string} that in the case of the Gross-Taylor string theory, it coincides precisely with the entanglement brane of \cite{Donnelly:2016jet}.}

In fact, the conditions \eqref{shrinkability} are not independent; they both follow from a new axiom. \\
\textbf{Entanglement brane axiom: }
\begin{equation} \label{ebraneaxiom}
 \qquad \mathtikz{\etaC{0cm}{0cm} } = \mathtikz{ \etaA{0cm}{0cm} \cozipper{0cm}{0cm}
\draw (0cm,0.5cm) node {\footnotesize $e$};
\draw (0cm,-0.25cm) node {\footnotesize $e$};
}.
\end{equation}
To see that this axiom implies \eqref{shrinkability}, we observe that:  
\begin{equation} \label{closewindow}
    \mathtikz{ \idC{0cm}{0cm} }
    = \mathtikz{ \idC{1.5cm}{-1cm} \pairC{0cm}{-1cm} \copairC{1cm}{-1cm} \idC{-0.5cm}{0cm} }
    = \mathtikz{ \epsilonA{0cm}{-2cm} \zipper{0cm}{-1cm} \idC{1.5cm}{-2cm} \idC{1.5cm}{-1cm} \idC{1.5cm}{0cm} \muC{0cm}{0cm} \copairC{1cm}{0cm} \idC{-0.5cm}{1cm} }
    = \mathtikz{ \epsilonA{0cm}{-2cm} \zipper{-0.5cm}{0cm} \zipper{0.5cm}{0cm} \idC{1.5cm}{-2cm} \idC{1.5cm}{-1cm} \idC{1.5cm}{0cm} \muA{0cm}{-1cm} \copairC{1cm}{0cm} \idC{-0.5cm}{1cm} }
    = \mathtikz{ \zipper{0cm}{1cm} \cozipper{0cm}{0cm} }
\end{equation}
This shows that closed string  ``windows" can be closed.  Moreover, we have:
\begin{equation}
\mathtikz{ \etaA{0cm}{0cm} }
= \mathtikz{ \zipper{0cm}{0cm} \etaC{0cm}{0cm} }
= \mathtikz{ \zipper{0cm}{0cm} \cozipper{0cm}{1cm} \etaA{0cm}{1cm} }
= \mathtikz{\muA{0}{-0.5cm} \tauA{0}{0.5cm} \deltaA{0}{1.5cm} \etaA{0cm}{1.5cm} }
= \mathtikz{ \copairA{0cm}{0cm} \muA{0cm}{0cm} },
\end{equation}
which implies
\begin{equation}
\mathtikz{ \idA{0cm}{0cm} }
= \mathtikz{ \muA{0cm}{0cm} \etaA{0.5cm}{0cm} }
= \mathtikz{ \muA{-0.5cm}{-1cm} \copairA{0cm}{0cm} \muA{0cm}{0cm} }
= \mathtikz{ \muA{-0.5cm}{-1cm} \copairA{0cm}{0cm} \muA{-1cm}{0cm} 
  \draw (-0.25cm,-1cm) to[out=90,in=-90] (0.25cm,0cm);
  \draw (0.25cm,-1cm) to[out=90,in=-90] (0.75cm,0cm);}
= \mathtikz{ \muA{0cm}{0cm} \deltaA{0cm}{1cm} }
\end{equation}
so both parts of Eq.~\eqref{shrinkability} is satisfied. 

Note that once we have the result \eqref{shrinkability}, any holes in the worldsheet can be closed up, so that any diagram with only closed inputs and outputs is equivalent to a diagram in the closed theory.
To see this, we use a result of \cite{Lauda:2005wn} which states that any diagram can be reduced to a normal form.
For diagrams without open inputs or outputs this normal form contains only closed cobordisms and some number of ``windows'' of the form
\begin{equation}
    i^*i = \mathtikz{ \cozipper{0cm}{0cm} \zipper{0cm}{1cm} }.
\end{equation}
Using \eqref{closewindow} we can close the windows, and we are left with a purely closed diagram.

This has important implications for the entanglement entropy.
Suppose we wish to calculate the entanglement entropy of an arbitrary state produced by the Euclidean path integral.
This state is represented by a cobordism from the empty set to some number of circles and intervals.
We can calculate the entanglement entropy by the replica trick, by calculating the partition function of the $\alpha$-replicated state as a function of $\alpha$.
For each integer $\alpha$, this partition function is a diagram without inputs or outputs and hence can be evaluated within the closed sector.
The result can be analytically continued in $\alpha$ to obtain the entanglement entropy.
Thus we see that we can evaluate the entanglement entropy purely within the closed sector, even though this entanglement is counting states within the open sector.
Thus the entanglement brane axiom implies that the closed sector carries information about the open sector.

Going in the reverse direction, the entanglement brane axiom also implies we can open up any closed diagram.
Since we can replace the cylinder with a window, we can also open up the product:\footnote{Note that this means that the closed product is determined by the open product.}
\begin{equation} \label{openproduct}
    \mathtikz{ \muC{0cm}{0cm} }
    = \mathtikz{ \cozipper{0cm}{0cm} \zipper{0cm}{1cm} \muC{0cm}{2cm} }
    = \mathtikz{ \cozipper{0cm}{0cm} \muA{0cm}{1cm} \zipper{-0.5cm}{2cm} \zipper{0.5cm}{2cm} }
\end{equation}
Now we can convert any closed diagram to open as follows.
First, we open up every unit and product using \eqref{openproduct} and the entangling brane axiom \eqref{ebraneaxiom}. 
The resulting diagram contains zipper/cozipper contractions of the form $ii^*$ which can be replaced with open diagrams using the Cardy axiom \eqref{axiom5}. 
The result is an equivalent diagram purely in the open sector.

Let us return now to the example of calculating the entanglement entropy of the Hartle-Hawking state \eqref{hhstate}.
Using the entanglement brane axiom, we can write the sphere diagram as a trace in the open sector, a relation which was essential to the original formulation of the entanglement brane in the string context \cite{Susskind:1994sm,Donnelly:2016jet}:
\begin{equation}
    \mathtikz{ \epsilonC{0cm}{0cm} \etaC{0cm}{0cm} }
    = \mathtikz{ \epsilonC{0cm}{-2cm} \zipper{0cm}{0cm} \cozipper{0cm}{-1cm} \etaC{0cm}{0cm} }
    = \mathtikz{ \epsilonA{0cm}{0cm} \etaA{0cm}{0cm} }
    = \mathtikz{ \epsilonA{0cm}{-1cm} \muA{0}{-0cm} \deltaA{0}{1cm} \etaA{0cm}{1cm} }
    = \mathtikz{ \pairA{0cm}{0cm} \copairA{0cm}{0cm} }.
\end{equation}
This shows that the modular flow associated with the Hartle-Hawking state, which is rotation on the sphere, can be instead expressed as a rotation on the annulus.
The latter can be interpreted as a trace, since the fixed points of the rotation can be replaced with a boundary satisfying the entanglement brane boundary condition.

In the next section we will see how this works in the specific example of Yang-Mills theory in two dimensions.

\section{Two-dimensional Yang-Mills theory}
\label{section:ym} 

The Euclidean partition function of  two-dimensional Yang-Mills theory on an arbitrary Riemann surface $M$ of area $A$ is given by
 \begin{align} \label{2dymz}
    Z= \sum_{R} (\dim R)^{\chi(M)} e^{-\frac{g^{2}_{\text{YM}} A C_{2} (R) }{2}} ,
 \end{align}
where $\chi(M)$ is the Euler characteristic and $R$ runs over irreducible representations of the gauge group $G$ (which is assumed to be compact).
Due to the dependence of \eqref{2dymz} on the area $A$, two-dimensional Yang-Mills is not purely topological except in the $A\rightarrow 0$ limit. 
However, the open-closed TQFT formalism can be easily extended to accommodate such an ``area-dependent" QFT \cite{Witten:1992xu,Cordes:1994fc,Runkel:2018uls}.
The main modification consists of attaching an area-dependent Boltzmann factor to each cobordism.
Because of the nontrivial Hamiltonian, the cylinder and strip cobordisms of nonzero area will now become propagators rather than the identity element.  
We will see that all the axioms of section \ref{section:tqft} and the entanglement brane axiom of \ref{section:ebrane} are satisfied with the only modification that total area must match on both sides of each formula.

\subsection{Two-dimensional Yang-Mills as an area-preserving QFT} 

Let us first consider the Hilbert space of a circle.
The configuration space variable is the holonomy $U=\text{P} \exp \left( \oint i A_\mu dx^\mu \right)$, and the corresponding Hilbert space consists of class functions of $U$.
A convenient orthonormal basis is given by states $\ket{R}$ whose wavefunctions are Wilson loops in the irreducible representations $R$:
 \begin{align}
     \braket{U|R}= \text{tr}_{R}(U).
 \end{align}
 In the zero-area limit, the unit element  which is compatible with the entanglment brane axiom has a wavefunction equal to the delta function on the group: 
 \begin{align}
     \braket{U|\eta_{C}} = \delta(U,1).
 \end{align}
This forces the holonomy along the boundary of each hole to be identity, so it can be shrunk down to a point.  
Expressed in the representation basis, this gives the state:
\begin{equation}
\eta_C = \sum_{R} \dim(R) e^{- \beta C_{2}(R) } \ket{R} = \mathtikz{ \etaC{0}{0} } \,.
\end{equation} 
Here we have introduced a dimensionless factor $\beta = \frac{1}{2} g_{YM}^2 A$ which acts like an inverse temperature.
On the interval, Gauss's law is relaxed at the boundaries and the Hilbert space is given by the space of square-integrable functions on $G$.  
Here, the orthonormal basis consists of matrix elements $\ket{Rab}$ in irreducible representations of $G$:
\begin{align}
    \braket{U|Rab} = \sqrt{\dim R} \; R_{ab}(U) \quad \quad a,b,=1 \cdots \dim R.
\end{align}
The factor of $\dim R$ ensures the states are normalized in the Haar measure.

The states labelled by $a,b$ are the edge modes that define the extension from the Hilbert space on a circle to that of an interval.
The entangling product which entangles these edge modes and glues together intervals corresponds to matrix multiplication.
This can be understood by expressing the Wilson line $U$ on the larger interval as a product $U=U_{V} U_{\bar{V}}$  of Wilson lines on the two halve of the interval. 
The wavefunction then factorizes according to 
\begin{align}
R_{ac}(U) &= \sum_{b} R_{ab}(U_{V}) R_{bc}(U_{\bar{V}} )  \nn
\ket{Rac} &\rightarrow \sum_{b}  \frac{1}{\sqrt{\dim R}} \ket{Rab}\ket{Rbc} 
\end{align} 
where the factor of $\frac{1}{\sqrt{\dim R}}$ accounts for the normalization of states.  This factorization defines the open multiplication 
\begin{align}
    \mu_\Open = \sum_{R,a,b,c} \frac{e^{- \beta C_{2}(R) }}{\sqrt{\dim(R)}} \ket{R a c} \bra{R a b} \bra{R b c}  = \mathtikz{ \muA{0}{0} }.
\end{align}
Similarly, a state on the circle can be embedded into the Hilbert space of an interval via:
\begin{align}
\text{tr}_{R}(U)&= \sum_{a} R_{aa}(U)\nn
\ket{R} &\rightarrow \sum_{a} \frac{1}{\sqrt{\dim R}}\ket{R aa},
\end{align} 
which defines the zipper.  
In summary, the rules that define 2D Yang Mills as an axiomatic QFT are:
\begin{align}\label{pp}
\mu_C &= \sum_{R} \frac{e^{-\beta C_{2}(R)}}{\dim(R)} \ket{R} \bra{R} \bra{R} &=& 
\mathtikz{ \muC{0cm}{0cm} } \\ 
\eta_C &= \sum_{R} \dim(R) e^{- \beta C_{2}(R)}\ket{R} &=& \mathtikz{ \etaC{0}{0} } \\
\mu_\Open &= \sum_{R,a,b,c} \frac{e^{- \beta C_{2}(R)}}{\sqrt{\dim(R)}} \ket{R a c} \bra{R a b} \bra{R b c}  &=& \mathtikz{ \muA{0}{0} } \\
\eta_\Open &= \sum_{R,a} \sqrt{\dim(R)} e^{-\beta C_{2}(R)}\ket{R a a} &=& \mathtikz{ \etaA{0}{0} } \\
i &= \sum_{R,a}  \frac{e^{-\beta C_{2}(R)}}{\sqrt{\dim(R)}} \ket{R a a} \bra{R} &=& \mathtikz{ \zipper{0}{0} }
\end{align}
Since the bases $\ket{R}$ and $\ket{Rab}$ are orthonormal, the corresponding co-units, co-multiplications, and co-zipper can be obtained simply by flipping bras to kets.

It can be verified that these assignments satisfy the Moore-Segal and entanglement brane axioms.
To see how the entanglement brane axiom is satisfied, it is useful to consider the annulus: 
\begin{align}
    \mathtikz{ \pairA{0cm}{0cm} \copairA{0cm}{0cm} } =  \sum_{R, a,b} e^{- \beta C_{2}(R)}= \sum_{R} (\dim R)^{2}e^{- \beta C_{2}(R)} =   \mathtikz{ \epsilonC{0cm}{0cm} \etaC{0cm}{0cm} } 
\end{align}
Each boundary of the annulus requires a trace over the edge modes supported there, giving a $\dim R$ factor per boundary.
This reproduces the sphere partition function and shows that the $(\dim R)^2$ factor in the closed sector counts edge modes in the open sector.

\subsection{Entanglement}

We now show how some explicit calculations of entanglement entropy can be carried out in two-dimensional Yang-Mills using the extended QFT formalism.

\paragraph{Single interval} 
As discussed in section \eqref{section:ebrane}, the factorization of the Hartle-Hawking state is given by the open copairing:
\begin{equation}
\mathtikz{ \deltaA{0cm}{0cm} \zipper{0cm}{1cm} \etaC{0cm}{1cm} }
= \mathtikz{ \deltaA{0cm}{0cm} \etaA{0cm}{0cm} } 
= \mathtikz{ \copairA{0cm}{0cm} }
= \sum_{R,a,b} e^{-\frac{1}{2} \beta C_{2}(R) }\ket{Rab} \ket{Rba}.
\end{equation}
where $\frac{\beta}{2}$ is the area of the hemisphere times $\frac{g_{YM}^2}{2}$.
Note that this state is unnormalized; as we will see shortly, the normalization factor is nonlocal.
Using the open pairing, we can turn this state on two intervals into a linear map from one to the other: 
\begin{equation}
\psi = \mathtikz{ \idA{-0.5cm}{0cm} \copairA{0cm}{0cm} \pairA{1cm}{0cm} \idA{1.5cm}{1cm} }=  \mathtikz{ \idA{0cm}{0cm}}= \sum_{R,a,b}   e^{-\frac{1}{2} \beta C_{2}(R)  }\ket{Rab} \bra{Rba}
\end{equation}
This state-channel duality is a useful trick. In particular we can write the (un-normalized) reduced density matrix on one interval as a strip of twice the area:
\begin{align}
    \rho = \psi \psi^{\dagger}=\mathtikz{ \idA{0cm}{0cm} \idA{0cm}{-1cm}}=\sum_{R,a,b}  e^{-\beta C_{2}(R) }\ket{Rab} \bra{Rba}
\end{align}
The modular Hamiltonian $H=-\log \rho$ generates evolution from one interval to the other; in this case it is given simply by $H = \beta C_{2}(R)$.  The trace of the density matrix defines a thermal partition function with respect to $H$:
\begin{align}
   Z= \text{tr} \rho = \sum_{R,a,b}  e^{-\beta C_{2}(R)  }  =   \mathtikz{ \pairA{0cm}{0cm} \copairA{0cm}{0cm} } 
\end{align}
which sums over edge modes propagating in each loop.
We can now read off the entanglement entropy from the eigenvalues of the reduced density matrix, which are $\frac{e^{-\beta C_{2}(R)}}{Z}$ (with multiplicity $\dim(R)^2$) when properly normalized.
In terms of the probability distribution over representations, $p(R)= \frac{(\dim R)^{2} e^{-\beta C_{2}(R)}}{Z} $, the entanglement entropy is 
\begin{align}\label{one}
    S= -\sum_{R} p(R) \log p(R)  + 2 \sum_{R}  p(R)
    \log \dim R
\end{align}
which agrees with the result in \cite{Donnelly:2014gva}.
The first term is the Shannon entropy of the distribution $p(R)$ which is associated with the sphere partition function.
The second counts the degeneracy of the edge modes, with the factor of $2$ corresponding to the two entangling points.

\paragraph{Two-intervals} 
The diagrammatic formalism generalizes easily for the case of multiple intervals.  For the case of two intervals, we can factorize the Hartle-Hawking state via: 
\begin{equation}
\mathtikz{ \deltaA{-1cm}{-1cm} \deltaA{1cm}{-1cm} \widedeltaA{0cm}{0cm} \zipper{0cm}{1cm} \etaC{0cm}{1cm} }
\end{equation} 
Using the state-channel mapping, can view this state as an evolution from one pair of intervals to the other
\begin{equation}\label{two}
\psi = 
\mathtikz{
\idA{0cm}{0cm} \muA{1.5cm}{0cm}
\idA{0cm}{1cm} \tauA{1.5cm}{1cm}
\deltaA{0.5cm}{2cm} \idA{2cm}{2cm}
}
= \sum_{R,a,b,c,d} \frac{e^{-\frac{1}{2}\beta C_{2}(R)} }{\dim R} \ket{Rab} \ket{Rcd} \bra{Rad} \bra{Rcb}
\end{equation}
This cobordism describes evolution under the modular Hamiltonian, and corresponds to half of the modular flow.  The reduced density matrix $\rho=\psi \psi^{\dagger} $ is obtained by flipping this diagram upside down and gluing it back to itself.   The effective partition function is now 
\begin{align}
    Z=\text{tr} \rho = \sum_{R,a,b,c,d} \frac{e^{-\beta C_{2} (R)}}{(\dim R)^{2} } \ket{Rab} \ket{Rcd} \bra{Rab} \bra{Rcd}
\end{align}
This is a path integral on a sphere with four holes.  Aside from the sum over more edges, this expression differs from the single interval case by the crucial $\frac{1}{(\dim R)^2}$ factor, which arises from the interaction of the two intervals.\footnote{ Due to this factor of $\dim(R)^{-2}$ and the sum over edge modes, the $n$-fold replica of the sphere will have the partition function $Z_{n} = (\dim R)^{4 - 2n} e^{- n \beta C_{2}(R)} $, which is consistent with $4-2n$ being the Euler characteristic of the replica manifold (for $n = 1$ it is a sphere, for $n=2$ a torus, etc.).} 
This modification of the Boltzmann factor is precisely what is needed to satisfy the entanglement brane axiom, since it makes $Z$ manifestly equal to the sphere partition function when we sum over all the edge modes. 
The two-interval modular hamiltonian is now
\begin{align}
     H = \sum_{R,a,b,c,d} \left( \log (\dim R)^2 + \beta C_{2}(R) \right) \ket{Rab}\ket{Rcd}\bra{Rab}\bra{Rcd}
\end{align}

The entanglement entropy can once again be expressed in terms of the probability distribution  $p(R)= \frac{(\dim R)^{2} e^{-\beta C_{2}(R)}}{Z} $ on the sphere: 
\begin{align}
   S= -\sum_{R} p(R) \log p(R)  + 4 \sum_{R}  p(R)
    \log \dim R
\end{align} 
This again splits in to a classical and a quantum piece which counts the edge mode degeneracy, with the factor of $4$ accounting for the boundaries introduced at four points of the entangling surface.

\paragraph{Thermal state}

Here we apply the extended TQFT formalism to decompose a thermal state.
The thermofield double state of two circles can be denoted by
\begin{equation}
    \ket{\text{TFD}} = \mathtikz{ \copairC{0cm}{0cm} }.
\end{equation}
We will consider the entanglement when one of these circles is split into two intervals:
\begin{equation}\label{TFD}
    \mathtikz{  \idC{-.5cm}{0 cm } \idC{-.5cm}{1cm } \zipper{1cm}{1cm} \widecopairC{0cm}{1cm} \deltaA{1cm}{0cm} } =
    \sum_{R,a,b} \frac{ e^{-\frac{1}{2} \beta C_2(R)}}{\dim R}  \ket{R} \ket{R,a,b}\ket{R,b,a}
\end{equation}

The single interval density matrix and the corresponding partition function are
\begin{align}
    \rho &= \mathtikz{ \muA{0cm}{-3cm} \zipper{0.5cm}{-2cm} \cozipper{0.5cm}{-1cm} \deltaA{0cm}{0cm}  \idA{-0.5cm}{-2cm} \idA{-0.5cm}{-1cm} } =\sum_{R,a,b} \frac{e^{-\beta C_{2} (R)}}{(\dim R)^2} \ket{Rab}\bra{Rab} \nn
    Z&=\text{tr}\rho= \sum_{R,a,b} \frac{e^{-\beta C_{2} (R)}}{(\dim R)^2}
\end{align}
The corresponding modular flow involves a open string that pinches to form a closed string and open string pair, which then recombine to form an open string.  
In this case, the normalization $Z=\sum_{R} e^{-\beta C_{2} (R)}$ is the partition function on a torus, as required by the E-brane axiom.
The entropy then takes the same form as in \eqref{one}, with $p(R)=\frac{ e^{-\beta C_{2} (R)}}{Z} $ the probability distribution on the torus.

\subsection{Reduced density matrix for general states and regions}

\begin{figure}[t]
\centering 
\includegraphics[scale=.4]{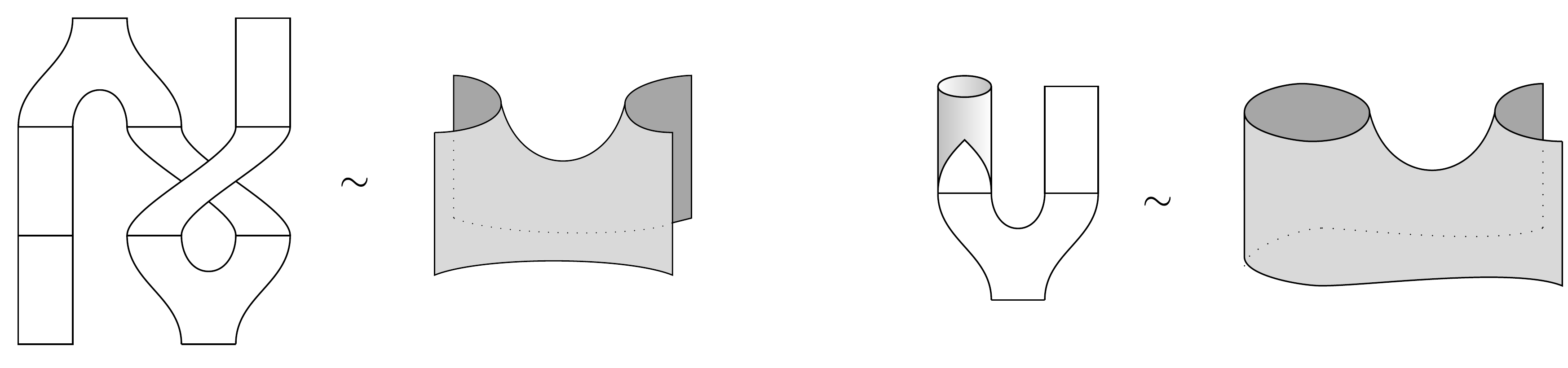}
\caption{These two cobordisms (along with the closed multiplication) describe saddle points of the modular flow.  
In these figures, modular time flows from top to bottom and can be interpreted as a height function, which is the standard example of a Morse function. 
A saddle point corresponds to a critical point of the Morse function where the Hessian has one positive and one negative eigenvalue.
Figure adapted from Ref.~\cite{Lauda:2005wn}.}
\label{2saddles}
\end{figure}

 Due to gauge invariance, the reduced density matrix of a general state and region is necessarily diagonal in the edge mode basis  $\ket{Rab}$ \cite{Donnelly:2011hn}.
 Therefore we only need to determine the correct powers of $\dim R$ that appear.
 In the examples above we saw that a factor of $\frac{1}{\dim R}$ is associated with the two cobordisms in figure \eqref{2saddles}.
 It was shown in \cite{Lauda:2005wn} that these correspond to saddle points of a Morse function, which we identify with the modular time parameter.
 
The occurrence of these saddle points is dictated by the global topology of the $2$-manifold and its foliation by intervals. 
In terms of the vector field that evolves these intervals in modular time, the saddle point is zero of index $-1$, while the entangling surface becomes  a zero of index $1$ when we shrink it to a point using the E-brane axiom\footnote{ The index of a zero is defined as the winding number of the map obtained by restricted the vector field to a circle around the zero.}.  We can then apply the Poincar\'e-Hopf theorem which says that the Euler characteristic $\chi$ of a manifold counts the total number of zeroes, graded by their index.
This implies that the number of saddle points that occur in an $n$-interval modular flow is 
\begin{equation}
    s= 2n-\chi.
\end{equation}

For example, applying this to the one interval foliation of a torus gives $s= 2-0=2$, in agreement with \eqref{TFD}.  For $n$ intervals on a sphere we have $s= 2n- 2$ saddle points; for example, when $n=3$ there should be four saddles.
The tensor product decomposition of this state is 
\begin{equation}
\mathtikz{
\draw (-0.75,-2) -- (-0.25,-2) to [out=-90,in=90,looseness=1] (-0.1,-3) -- (-0.6,-3) to [out=90,in=-90,looseness=1] (-0.75,-2);
\draw (0.25,-2) -- (0.75,-2) to [out=-90,in=90,looseness=1] (0.6,-3) -- (0.1,-3) to [out=90,in=-90,looseness=1] (0.25,-2);
\deltaA{1.5cm}{-2cm}\deltaA{-1.5cm}{-2cm}\deltaA{-1cm}{-1cm} \deltaA{1cm}{-1cm} \widedeltaA{0cm}{0cm} \zipper{0cm}{1cm} \etaC{0cm}{1cm} }.
\end{equation} 
The cobordism describing half the modular flow gives the linear mapping
\begin{equation}
    \psi= \sum_{R,a,b,c,d,e,f}\frac{e^{-\frac{1}{2} \beta C_{2}(R)} }{(\dim R)^2} \ket{Rad}\ket{Rcf}\ket{Reb} \bra{Rab}\bra{Rcd}\bra{Ref} 
\end{equation}
so $\rho= \frac{e^{-\beta \hat{C}_{2}(R)}}{(\dim R)^4} $, consistent with the presence of four saddle points.

\subsection{Negativity}
\label{section:negativity}

While the entanglement entropy is a useful characterization of entanglement for pure states, for a mixed state it does not distinguish between entanglement and classical correlations.
For such states more refined measures of entanglement exist, but unfortunately most are not easily computable.\footnote{A precise statement of their noncomputability is given in Ref.~\cite{Huang_2014}, and we thank Yichen Huang for bringing this work to our attention. Of course, this does not preclude the possibility of computing it for specific states, such as those prepared by the Euclidean path integral as we considered here.}
One exception is the logarithmic negativity, which we will now consider \cite{PhysRevA.65.032314,Plenio:2005cwa}.

For a density matrix $\rho$ on two intervals, we define its partial transpose $\rho^\Gamma$ by:
\begin{equation}
\mathtikz{
\idA{-2.5cm}{1cm};
\idA{-1.5cm}{1cm};
\idA{-2.5cm}{-1cm};
\idA{-1.5cm}{-1cm};
\draw[thick] (-3cm,-1cm) rectangle node{$\rho^\Gamma$} (-1cm,0cm);
}
= 
\mathtikz{ \tauA{0cm}{0cm}; \pairA{-1cm}{-1cm}; \copairA{-1cm}{0cm}; 
\idA{-2.5cm}{1cm};
\idA{0.5cm}{1cm};
\idA{-2.5cm}{-1cm};
\idA{0.5cm}{-1cm};
\draw[thick] (-3cm,-1cm) rectangle node{$\rho$} (-1cm,0cm);
}
\end{equation}
Note that in general one must be careful about complex conjugation so that one takes the transpose and not the hermitian conjugate; in this example all wavefunctions are real and we can ignore this subtlety.
Partial transposes of more general states can be similarly expressed by flipping some subset of inputs and outputs using the pairing and copairing.

While $\rho$ defines a positive operator, $\rho^\Gamma$ is not necessarily positive.
The logarithmic negativity
\begin{equation}
    \mathcal{E} = \log \lVert \rho^\Gamma \rVert_1,
\end{equation}
measures the failure of $\rho^\Gamma$ to be positive and acts as a useful measure of entanglement.

As an example, consider negativity of two adjacent intervals on the sphere, which share a single endpoint.
The reduced density matrix in this situation is given by:
\begin{equation}
\rho = \mathtikz{ \deltaA{0cm}{0cm} \muA{0cm}{1cm} }.
\end{equation}
The partial transpose of the right interval is: \begin{equation} \label{rhogamma}
    \rho^\Gamma = \mathtikz{
\idA{0cm}{0cm} \muA{1.5cm}{0cm}
\idA{0cm}{1cm} \tauA{1.5cm}{1cm}
\deltaA{0.5cm}{2cm} \idA{2cm}{2cm}
} = \sum_{R} \frac{e^{-\beta C_{2}(R)} }{\dim R} \ket{Rab} \ket{Rcd} \bra{Rad} \bra{Rcb}.
\end{equation}
Note that if we trace any odd power of $\rho^\Gamma$ we get a surface with three boundaries, while if we trace any even power of $\rho^\Gamma$ we get a surface with four boundaries.

We can calculate the negativity by separately analytically continuing from odd and even $n$ \cite{Calabrese:2012ew}.
Let $Z_{n_e} = \text{tr}((\rho^\Gamma)^{n_e})$ for $n_e$ even and $Z_{n_o} = \text{tr}((\rho^\Gamma)^{n_o})$ for $n_o$ odd.
For the case at hand we have
\begin{equation}
    Z_{n_e} = \sum_{R} \dim(R)^{4-n_e} e^{-n_e \beta C_2(R)}, \qquad
    Z_{n_o} = \sum_{R} \dim(R)^{3-n_o} e^{-n_o \beta C_2(R)}.
\end{equation}
$n_o \to 1$ just gives the normalization of the state.
The logarithmic negativity is given by
\begin{equation}
    \mathcal{E} = \lim_{n_e = n_o \to 1} \log \frac{Z_{n_e}}{Z_{n_o}}.
\end{equation}
Note that this formula does not require $Z$ to be normalized.

We can write this in terms of the distribution over representations, which is
\begin{equation}
    p(R) = \frac{1}{Z_1} \dim(R)^2 e^{-\beta C_2(R)}.
\end{equation}
And we find that
\begin{equation}
    \mathcal{E} = \log \langle \dim (R) \rangle.
\end{equation}
We can find the same result by calculating the spectrum of $\rho^\Gamma$ given by \eqref{rhogamma}.
Note that this is \emph{not} the same as the edge term $\langle \log \dim R \rangle$.
Instead we have $\langle \log \dim R \rangle < \log \langle \dim R \rangle$ by convexity of the logarithm.
We will comment on this distinction in the next subsection.

We can also consider the negativity of two adjacent intervals in a thermofield double state \eqref{TFD}.
The partially transposed density matrix is 
\begin{equation}
\rho^{\Gamma}= \mathtikz{ \idA{0cm}{0cm}  \idA{0cm}{-1cm} 
\semiwidedeltaA{1.75cm}{0cm}
\muA{2.5cm}{-3 cm} \tauA{2.5cm}{-2cm} \deltaA{2.5cm}{-1cm} \idA{1cm}{-1cm} \tauA{.5cm}{-2cm}  \idA{1cm}{-3cm}   \semiwidemuA{1.75cm}{-5cm}  \idA{0cm}{-3cm}
\idA{0cm}{-4cm}  } = \sum_{R,a,b,c,d}\frac{e^{-\beta C_{2}(R)}}{(\dim R)^2} \ket{Rab}\ket{Rcd}\bra{Rcd}\bra{Rab} 
\end{equation}
Apart from  the factor of $\frac{1}{(\dim R)^{2} }$, the density matrix just swaps the state on the two intervals.
The normalization is the torus partition function 
\begin{equation}
   Z = \text{Tr} \rho^{\Gamma} = \sum_{R ,a,b,c,d} \frac{e^{- \beta C_{2}(R)}}{(\dim R)^{2} }\delta_{ac}\delta_{bd} = \sum_{R} e^{-\beta C_{2}(R)}
\end{equation}
  An eigenbasis of  $\rho^{\Gamma}$ is given by 
 \begin{equation}
     \ket{Rabcd \pm} = \frac{1}{\sqrt{2}} \left(\ket{Rab}\ket{Rcd}\pm \ket{Rcd}\ket{Rab}\right)
 \end{equation}
 where the antisymmetric state is absent when $a=c,b=d$.  
 The normalized eigenvalues are $ \pm \frac{e^{-\beta C_{2}(R)}}{(\dim R)^{2} Z }$
 The negativity is therefore 
 \begin{equation}
    \mathcal{E} = \log \sum_{R,a,b,c,d} \frac{e^{-\beta C_{2}(R)}}{(\dim R)^{2} Z  }  = \log \sum_{R} \frac{(\dim R)^{2} e^{-\beta C_{2}(R)}}{ Z  }= \log \braket{(\dim R)^{2}},
 \end{equation}
where the expectation value is taken with respect to the probability distribution $p(R)=\frac{e^{-\beta C_{2}(R)}}{Z}$ on the torus.

We see that the logarithmic negativity captures the entanglement in the edge modes between adjacent intervals, which is associated with the powers of $\dim(R)$.
Unlike the von Neumann entropy, it does not pick up the Shannon entropy associated with the distribution over different representations $R$.
This reflects the structure of the states: the label $R$ is a global degree of freedom, so when we reduce to a subregion it is effectively a classical degree of freedom.
Conversely, we see that the edge modes do contribute to the negativity and correspond to entanglement rather than simply classical correlations.\footnote{It was noted in \cite{Soni:2015yga,VanAcoleyen:2015ccp} that the entanglement associated with the edge modes cannot be distilled into Bell pairs with gauge-invariant operations. While this is true, the logarithmic negativity is blind to the distinction between gauge-invariant and gauge-variant operators and so counts the edge modes as entangled.}

\subsubsection{Relative negativity?}

The logarithmic negativity bears some similarity to the edge mode contribution to the entanglement, except that it takes the form of $\log \langle \dim(R)^n \rangle$ rather than $\langle \log \dim(R)^n \rangle$.
In this sense the logarithmic negativity is more analogous to a free energy than to an entropy.
However it is suggestive of a related measure that would give the edge term exactly.

We will consider again the example of two adjacent intervals on the sphere, for which the reduced density matrix (normalized) is
\begin{equation}
\rho = \frac{1}{Z} \sum_{R,a,b,c,d} \frac{e^{-\beta C_2(R)}}{\dim(R)} \ket{Rab} \ket{Rbc} \bra{Rad} \bra{Rdc},
\end{equation}
and its partial transpose is
\begin{equation}
\rho^\Gamma = \frac{1}{Z} \sum_{R,a,b,c,d} \frac{e^{-\beta C_2(R)}}{\dim(R)} \ket{Rab} \ket{Rdc} \bra{Rad} \bra{Rbc}.
\end{equation}

Then a natural quantity to consider is the relative entropy between $\rho$ and $\rho^\Gamma$:
\begin{equation} \label{relativenegativity}
S(\rho^\Gamma || \rho) = \tr(\rho \log \rho) - \tr(\rho \log \rho^\Gamma).
\end{equation}
Note that $\log(\rho^\Gamma)$ would appear to be ill-defined, since $\rho^\Gamma$ can have negative or zero eigenvalues.
In the present case we see that the negative eigenspace of $\rho^\Gamma$ is spanned by antisymmetric states of the form $\ket{R ab}\ket{Rdc} - \ket{Rad}\ket{Rbc}$, on which $\rho$ has no support.
A straightforward calculation then gives for this example
\begin{equation}
S(\rho^\Gamma || \rho) = \sum_R p(R) \log \dim(R) = \langle \log \dim(R) \rangle.
\end{equation}
Thus the relative entropy captures precisely the entropy of the edge mode shared by the two intervals.

Unfortunately, the quantity \eqref{relativenegativity} does not make sense in general; the operators $\rho$ and $\rho^\Gamma$ act on different spaces.
Nevertheless, it might be interesting to try to find an analog of \eqref{relativenegativity} that would be well-defined for more general quantum systems.

\section{The Gross-Taylor string theory}
\label{section:string}

The large-$N$ limit of two-dimensional Yang-Mills theory can be formulated as a closed string theory \cite{Gross:1993hu}. 
In particular, the $U(N)$ Yang Mills partition function on a closed Riemann surface admits a closed worldsheet expansion with string coupling $g_{\text{string}}=\frac{1}{N}$.  
In addition to the usual string interactions, it was found that the string encounters certain target space singularities called $\Omega$ and $\Omega^{-1}$ points, whose presence depends on the target space topology.   
In \cite{Donnelly:2016jet}, we considered the partition function on a sphere as an effective thermal partition function describing entanglement between two halves of the equatorial circle.
We showed that the two $\Omega$ points on the sphere are the entangling points cutting the circle into two intervals.
These entangling points can be stretched into the worldline of an entanglement brane, which is a hypersurface where open strings end. 
Unlike a D-brane, the open strings ending on an entanglement brane are part of a closed string which is partially hidden behind the entangling surface. 

To show that we can consistently treat the entanglement edge modes of the Gross-Taylor string theory in terms of entanglement branes, we have to consider entanglement of multiple disjoint intervals.
The corresponding modular flow probes Riemann surfaces of negative Euler characteristic, where we can give an open string interpretation of the $\Omega^{-1}$ points.  
Given these motivations, we now apply the extended TQFT formalism to the target space of the Gross-Taylor string.      

\subsection{Chiral Gross-Taylor string as a closed string TQFT}
 
 The Hilbert space of the Gross-Taylor string can be obtained as a certain large-$N$ limit of the Yang-Mills Hilbert space.  
 Since the Yang-Mills Hilbert space is labelled by representations of $U(N)$, a prescription is needed for how to fix a representation while taking $N \rightarrow  \infty$.
 The naive way of taking this limit via the Frobenius formula, in which one keeps the number of boxes in the Young tableau finite as $N \to \infty$, captures one chiral half of the Yang Mills Hilbert space \cite{Gross:1993yt}.
 In the following we consider the corresponding chiral Gross-Taylor (CGT) string theory. 
 We will use the extended TQFT framework to formulate this string theory independently from its gauge theory origins.
 
 We consider the chiral case purely for simplicity; the non-chiral theory can also be treated in this axiomatic framework but the precise form of the entanglement brane is more complicated \cite{Donnelly:2016jet}.

\paragraph{Hilbert space on a circle}

The (second quantized) Hilbert space of closed strings is isomorphic to the fock space of bosonic oscillators $a_{l}^{\dagger}$ which create strings winding $l$ times.  String configurations with total winding number $n$ are described by states labelled by permutations $\sigma \in S_{n}$.
\begin{align}\label{CS} 
|\sigma \rangle  =  \prod_{l=1}^{\infty} (a_{l}^{\dagger} )^{n_{l}}\ket{0}, \quad \quad \quad  \sum{l} n_{l} =n 
\end{align}
where $n_{l}$ denotes the number of cycles of length $l$ in $\sigma$.  
Each cycle represents a closed string loop that winds $l$ times.  
Closed string are indistinguishabe, and this is reflected in the the fact that the state $\ket{\sigma}$ only depends on the conjugacy class of $\sigma$ as specified by $n_{l}$.
The closed string inner product follows from the standard commutation relations 
\begin{align}
 [a^{\phantom{\dagger}}_{l},a^{\dagger}_{l'}] &= l \, \delta_{l l'}
 \end{align}
In terms of permutations, we have 
\begin{equation}
    \braket{\sigma|\eta} = \sum_{\tau} \delta(\eta, \tau \rho \tau^{-1})
\end{equation}
with the corresponding resolution of identity
\begin{align}
   \mathbf{1}= \sum_{n} \sum_{\sigma \in S_{n} } \frac{1}{n!} \ket{\sigma}\bra{\sigma}  =\mathtikz{\idC{0cm}{0cm}}
\end{align}
Note that the states $\ket{\sigma}$ are overcomplete and also not normalized.

\paragraph{Basic cobordisms}

For convenience we will consider the non-interacting limit, which is the analogue of the zero area limit of 2D Yang-Mills.\footnote{Restoration of the string interactions is slightly more complicated than the case of Yang-Mills, because the string basis does not diagonalize the Hamiltonian.  As shown in \cite{Donnelly:2016jet}, this involves adding branch point interactions in the bulk of the string worldsheets and summing over their locations in target space, resulting in an area dependence.}
The basic cobordisms that define the CGT string theory at closed string coupling $g_{s}$ are:
\begin{align}
\eta_\C &= \sum_{n=1} \frac{1}{n!} \sum_{\sigma \in S_{n}} g_{s}^{-K_{\sigma}}  \ket{\sigma } &=& \mathtikz{ \etaC{0}{0} } \\
\mu_\C(\ket{\sigma}\ket{\tau}) &= 
\sum_{\rho \in S_n} g_{s}^{n}\omega_{ \sigma \tau \rho } \ket{\rho}
&=& 
\mathtikz{ \muC{0cm}{0cm} } \\
\epsilon_\C \ket{\sigma} &= g_{s}^{-K_{\sigma}} &=& \mathtikz{ \epsilonC{0}{0} } \\
\pi_\C( \ket{\sigma} \ket{\eta} ) &= \sum_{\tau} \delta(\eta, \tau \sigma \tau^{-1}) &=& \mathtikz{ \pairC{0cm}{0cm}} \\
\kappa_\C &= \sum_n \frac{1}{n!} \sum_{\sigma \in S_n} \ket{\sigma} \ket{\sigma} &=& \mathtikz{ \copairC{0cm}{0cm}}
\end{align}
Here we have introduced new notation for the pairing $\pi$ and co-pairing $\kappa$; because the states $\ket{\sigma}$ do not form an orthonormal basis, these have a nontrivial representation.
For the same reason, we have expressed the cobordisms by their action on states rather than using bra-ket notation which uses the inner product implicitly.
The factors $\omega_\sigma$ appearing in the product will be determined below.

Each cobordism defines a target space for closed, string worldsheets which wrap the spacetime at least once.  The closed string unit $\eta_{\C}$ is the fundamental building block for the closed TQFT.  Since it is a contractable hemisphere, we demand that the connected components of the allowed worldsheets have disk topologies and end on the $S^{1}$ bounding the hemisphere.  We assign the state $\ket{\sigma}$ to each worldsheet whose boundary covers the $S^{1}$ according to the permutation $\sigma$.  The corresponding amplitude then assigns a factor of $g^{-1}_{s}$ for each disk that appears in the worldsheet configuration, consistent with the rules of string perturbation theory. 

We can think of the center of the hemisphere as the location of a D-instanton that emits closed strings with amplitude  $g_{s}^{-1}$ per closed string \cite{Polchinski:1994fq}. 
This is also a branch point for the closed string worldsheets, whose branches are permuted according to $\sigma$ as we encircle the D-instanton.
This is what Gross and Taylor referred to as the $\Omega$ point.

We now turn to the definition of the weights $\omega_\sigma$. 
The multiplication $\mu_{C}$ can be written generically as
\begin{align}
\mu_\C(\ket{\sigma}\ket{\tau}) &= 
\sum_{\rho \in S_n} g_{s}^{n}\omega_{ \sigma \tau \rho } \ket{\rho}
\mathtikz{ \muC{0cm}{0cm} }
\end{align}
where each term corresponds to worldsheet process in which initial closed string configurations $\sigma$ and $\tau$ are cut and reglued into the final configuration $\rho$.
The factor of $g_{s}^n$ has been factored out of the weight $\omega$ for convenience. 
To obtain the correct fusion amplitudes we apply the constraint 
\begin{align}\label{mu}
    \mathtikz{ \muC{.5cm}{1cm}
    \etaC{0cm}{ 1.cm}} = \mathtikz{\idC{0cm}{0cm}},
\end{align}
which identifies $\mu_{C}$ as the multiplicative inverse of $\eta_{C}$.  
Then for each $n$, the coefficients $\omega_{\alpha}$ are determined by first defining an element  $\Omega_{n}$ of the $S_{n}$ group algebra: 
\begin{align}
    \Omega_{n} = \sum_{\alpha \in S_{n}} g_{s}^{n-K_{\alpha}} \alpha 
\end{align}  
and then taking the formal inverse:
\begin{align}
\Omega_{n}^{-1} =\sum_{\alpha \in S_{n} } \omega_{\alpha } \alpha 
\end{align}
 The identity $\Omega \Omega^{-1} = \mathbf{1}$ in the $S_n$ algebra then implies
 \begin{equation}
    \sum_{\mathbf{\alpha} \in S_{n} } (g_s^{n -K_{\alpha}}  \omega_{\alpha \sigma^{-1} }) = \delta (\sigma),
\end{equation}
 which ensures equation \eqref{mu} is satisfied.
 
As in the case of the hemisphere, the worldsheets form a branched covering of the diagram representing $\mu_\C$, which is sphere with three punctures.  
The punctures are labelled by string configurations $\tau$, $\rho$, and $\sigma$ which determine the branching structure of the worldsheet around those points. 
Fusing these branch points together gives a branch point singularity that is weighted by $\omega_{\tau \rho \sigma }$.
This is what Gross and Taylor called an $\Omega^{-1}$ point.
The difference between the number of $\Omega$ points and the number of $\Omega^{-1}$ points is equal to the Euler characteristic of the target space.

\subsection{Chiral Gross-Taylor string as an open-closed TQFT}
\paragraph{Hilbert space on an interval}

Upon cutting the CGT closed string into an open string, one finds that each open string endpoint supports $N$ edge modes corresponding to Chan-Paton indices of $U(N)$. 
Formally, the CGT string assigns to an interval the Hilbert space of functions on the group $U(N)$ as $N \rightarrow \infty$.  
This open string Hilbert space is spanned by  wavefunctions of the form
 \begin{align}
     \braket{U|IJ} = U_{i_{1}j_{1}} \dots U_{i_{n}j_{n}}, 
 \end{align}
 where $U_{ij}$ is a matrix element in the fundamental representation and $n > 0$ counts the number of open strings. 
 The multi-dimensional Chan-Paton factors $I=(i_{1},\dots,i_{n})$ and $J=(j_{1},\dots,j_{n})$ actually give a redundant labelling of the states since $\ket{IJ} =\ket{\sigma(I)\sigma(J)}$ for any permutation $\sigma \in S_{n}$.  This is an expression of open string indistinguishability. 
 Moreover these states are not orthogonal, instead their inner product is given by 
 \begin{align}\label{oi}
    \mathtikz{\pairA{0cm}{0cm} }=\int d U \,\,  U_{IJ} U_{KL}^{\dagger} = \sum_{\alpha, \sigma \in S_{n}} \frac{\omega_{\alpha}}{N^n} \delta_{I, \sigma(L) } \delta_{J,  \alpha \sigma(K) }.
\end{align}
 This formula says that the overlap between two stacks of open strings is zero unless we can tie their Chan-Paton factors together to make a closed string configuration. As in the closed string case, the coefficients $\omega_{\alpha}$ are defined via 
\begin{align}
    \Omega_{n} = \sum_{\alpha \in S_{n}} N^{K_{\alpha}-n} \alpha,\quad \quad  \Omega_{n}^{-1} = \sum_{\alpha \in S_{n}} \omega_{\alpha} \alpha
\end{align}
The appearance of $\Omega^{-1}_{n}$ can also be understood via it's relation to the dimensions $\dim(R)$ of $U(N)$ irreps.  
Using Schur-Weyl duality, it can be shown that for an irreducible representation $R$ corresponding to a Young tableau with $n$ boxes 
\begin{align}
(\dim R)^{\pm 1}  =  \frac{1}{n!} \chi_{R}(\Omega^{\pm 1}_{n}) ,
\end{align}
where $\chi_{R}(\alpha)$ is a character of $S_{n}$.
Due to the grand orthogonality theorem, the $U(N)$ inner product evaluated in the representation basis will contain a factor of  $(\dim R)^{-1}$  which is responsible for the appearance of $\Omega^{-1}$ in the open string formula \eqref{oi}.
The non-orthogonal inner product implies a non-trivial isomorphism between the Hilbert space and its dual, which makes the usual bra-ket notation for linear maps problematic.
For this reason we will avoid this notation in the following.

\paragraph {The basic open-closed cobordisms}

The open cobordisms describe target spaces for open string worldsheets.  The  ``constrained boundaries" traced out by endpoints of intervals are worldlines of 0-branes where open strings end.  This gives a worldsheet description of the entanglement brane. 

Here we outline a way to systematically build up the open string extension of the closed string algebra, starting with the co-multiplication $\Delta_{\Open}$. 
This describes the extended Hilbert space factorization which splits each initial open string to into two according to  $U_{ij}=\sum_{k} U_{ik}U_{kj}$:
\begin{align}
    \Delta_\Open(\ket{IJ}) &= \sum_K \ket{IK} \ket{KJ} = \mathtikz{ \deltaA{0cm}{0cm} }.
\end{align}
As in the case of Yang-Mills the sum over $k$ projects the internal edge modes onto a singlet under $U(N)$.
The same factorization can be applied to a closed string state $\ket{\sigma }$, viewed as an element of the open string Hilbert space.
This leads to the expression for the zipper $i$.
\begin{align}
i(\ket{\sigma}) = \sum_{I} \ket{I \sigma(I)}
 = \mathtikz{ \zipper{0}{0} }
\end{align}  From the co-multiplication, we can also obtain the co-unit $\epsilon_{\Open}$ via the identity
\begin{align}
\mathtikz{   \epsilonA{-.5cm}{-1 cm} \deltaA{0}{0}}=\mathtikz{\idA{0}{0}} \quad \Rightarrow \quad \epsilon_{\Open}(\ket{IJ}) =\delta_{IJ}.
\end{align}

Once equipped with the zipper $i$, we can apply axiom 1 \eqref{axiom1} to obtain the open string unit from the closed string unit: 
 \begin{align}
     \eta_{A} = i \circ \eta_{\C} =  \sum_{n=1}\sum_{I} \frac{1}{n!} \sum_{\sigma \in S_{n} } g_{s}^{-K_{\sigma}}  \ket{I\sigma(I)}.
 \end{align}
 This expression contains $g_{s}$, which requires an open string interpretation. This is determined by the entanglement brane axiom in the form.
 \begin{align}
  \mathtikz{ \zipper{0}{0} \epsilonA{0cm}{-1 cm}}= \mathtikz{\epsilonC{0}{0}}
 \end{align} 
Applying this equation to a basis state $\ket{\sigma}$ gives $  \sum_{I} \delta_{I, \sigma(I)}= N^{K_{\sigma}} =g_{s}^{-K_{\sigma}}$, leading to the crucial relation
\begin{align}
   g_{s} = \frac{1}{N} 
\end{align}
This is a compatibility relation needed in order to shrink a D$0$ brane into a D-instanton, in accordance with the entanglement brane axiom.

The unit can now be combined with the co-multiplication to give the co-pairing: 
\begin{align}
    \mathtikz{\copairA{0cm}{0cm}} = \mathtikz{ \deltaA{0cm}{0cm} \etaA{0cm}{0cm}} = \sum_{n=1}\sum_{I} \frac{1}{n!} \sum_{\sigma \in S_{n} } N^{K_{\sigma}}  \ket{IK}\ket{K\sigma(I)}.
\end{align}
As in the closed string TQFT, the open string multiplication is determined by taking the multiplicative inverse of the open unit, which leads to the appearance of the $\Omega^{-1}$ point.  Alternatively, it can be determined by the gluing:
\begin{align}
\mathtikz{
 \muA{0cm}{0cm}} = \mathtikz{\deltaA{0cm}{0cm} \pairA{1cm}{-1cm} \idA{1.5cm}{0cm} \idA{-0.5cm}{-1 cm}} \quad \Rightarrow \quad \mu_\Open(\ket{IJ}\ket{KL}) &=\sum_{\alpha,\sigma \in S_{n}}\frac{\omega_{\alpha \sigma^{-1}}}{N^n} \delta_{J,\alpha (K)} \ket{I \sigma(L)}
 \end{align}
The corresponding worldsheets describe the fusion of open strings whose Chan-Paton indices are matched up to permutations.

Finally, the co-zipper can be obtained by the gluing 
\begin{align}
    \mathtikz{ \cozipper{0cm}{1cm}  } =  \mathtikz { \pairA{0cm}{0cm} \zipper{-0.5cm}{1cm}  \idC{-1.5cm}{1cm} \copairC{-1cm}{1cm} \idA{.5cm}{1cm}}
\end{align}
and the pairing from solving the zigzag identity:
\begin{equation}
\mathtikz{\copairA{0cm}{0cm} \pairA{1cm}{0cm}}=  \mathtikz{ \idA{0cm}{0cm}}
\end{equation}
We leave some details of the basic cobordism calculations to the appendix.  In summary, the elementary open-closed cobordisms are completed by the following: 

\begin{align}
\mu_\Open(\ket{IJ}\ket{KL}) &=\sum_{\alpha,\sigma \in S_{n}}\frac{\omega_{\alpha \sigma^{-1}}}{N^n} \delta_{J,\alpha (K)} \ket{I \sigma(L)}   &=& \mathtikz{ \muA{0}{0} } \\
\Delta_\Open(\ket{IJ}) &= \sum_K \ket{IK} \ket{KJ} &=& \mathtikz{ \deltaA{0cm}{0cm} } \\
\eta_\Open &=  \sum_{n=1}\sum_{I} \frac{1}{n!} \sum_{\sigma \in S_{n} } N^{K_{\sigma}}  \ket{I\sigma(I)} &=& \mathtikz{ \etaA{0}{0} } \\
\epsilon_{\Open}(\ket{IJ})&= \delta_{IJ}&=& \mathtikz{ \epsilonA{0}{0} } \\
i( \ket{\sigma} ) &= \sum_{I}\ket{I\sigma(I)} &=& \mathtikz{ \zipper{0}{0} } \\
i^*(\ket{IJ}) &= \sum_{\tau, \sigma} \frac{\omega_\sigma}{N^n} \delta(J, \sigma \tau^{-1} I) \ket{\sigma} &=& \mathtikz{ \cozipper{0}{0} }
\end{align}

\paragraph{Examples}
Here we apply the basic cobordisms described above to some entanglement calculations and provide a worldsheet interpretation that incorporates the $\Omega$ points and $\Omega^{-1} $ points first observed by Gross and Taylor \cite{Gross:1993yt}. 

We begin by reviewing the result of \cite{Donnelly:2016jet}, which gave a worldsheet description of the entanglement between two intervals in the state given by the hemisphere.  As noted in section \eqref{section:ebrane} the cobordism describing the tensor factorization of this state is
\begin{align}
\ket{\psi}&= \mathtikz{ \deltaA{0cm}{0cm} \zipper{0cm}{1cm} \etaC{0cm}{1cm} }
= \mathtikz{ \deltaA{0cm}{0cm} \etaA{0cm}{0cm} } 
= \mathtikz{ \copairA{0cm}{0cm} }
\end{align}
The state reduced to one interval has a density matrix whose trace is given by
\begin{align}
     \mathtikz{ \pairA{0cm}{0cm} \copairA{0cm}{0cm} }= \text{tr} \left( \mathtikz{\idA{0}{0}} \right) =
    \sum_{n}\sum_{IJ} 1 = \sum_{n} \sum_{\sigma \in S_{n}}\frac{ N^{2 K_{\sigma}}}{n!} 
\end{align}
In the last equality, we accounted for open string distinguishability. This is a thermal partition function where each term describes  disconnected open strings worldsheet ending on N entanglement branes at each boundary. The permutation $\sigma$ determines the winding of the open strings around the thermal circle, and $K_{\sigma}$ counts the number of open strings. Applying the entanglement brane axiom gives a closed string description: 
\begin{align}
   \mathtikz{ \pairA{0cm}{0cm} \copairA{0cm}{0cm} } =\mathtikz{ \epsilonC{0cm}{0cm} \etaC{0cm}{0cm} }=\sum_{\sigma \in S_{n}} \frac{g_{s}^{-2 K_{\sigma}}}{
   n!} 
\end{align} 
Here each stack of $N$ entanglement branes is shrunk to an $\Omega$ point.
Each term in the sum corresponds to a covering of the sphere by a (disconnected) closed worldsheet that is branched over the two $\Omega$ points. 
The permutation $\sigma \in S_{n}$ labels the pattern of branching and determines the worldsheet Euler characteristic, which in turn determines the powers of $g_{s}$.  
The number of $\Omega$ points coincides with the Euler characteristic $\chi=2$ of the target space, consistent with the result of \cite{Gross:1993hu,Gross:1993yt}. 

To identify general entangling surfaces with E-branes and the $\Omega$ point singularities, we consider the density matrix for multiple intervals in the Hartle Hawking state. 
For two intervals, we learned in section \ref{section:ym} that half of the modular flow is described by the cobordism
in \eqref{two}. 
\begin{figure}[h]
\centering 
\includegraphics[scale=.4]{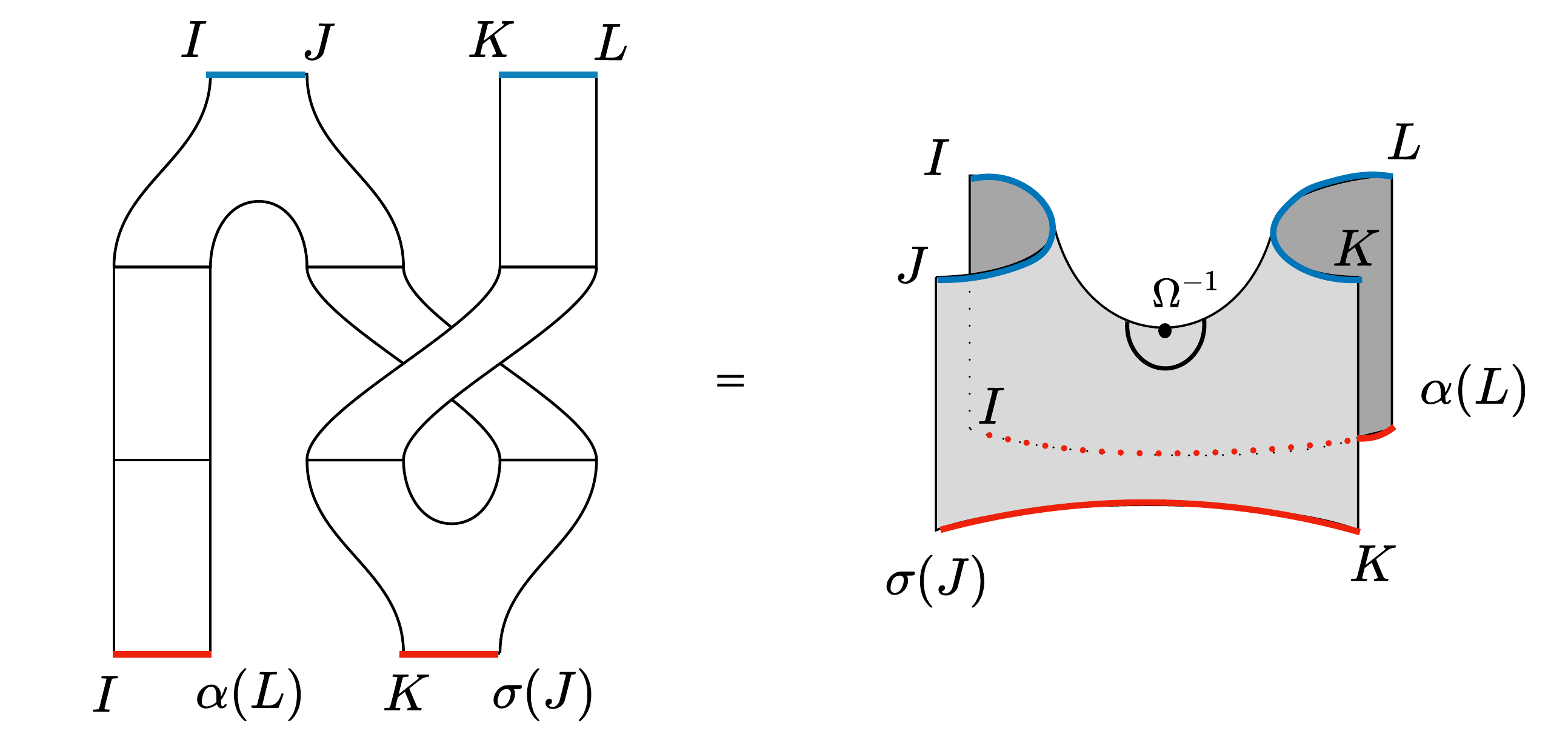}
\caption{The cobordism associated with half of the modular flow of two intervals on $S^2$, and an associated Morse function with one singularity. Figure adapted from Ref.~\cite{Lauda:2005wn}.}
\label{saddle}
\end{figure}

In terms of formulas, this gives a mapping
\begin{align}
    \psi(\ket{IJ} \ket{KL}) = \sum_{\alpha,\sigma \in S_{n}} \frac{\omega_{\sigma \alpha}}{N^{n}} \ket{I \alpha(L)} \ket{k \sigma(J)}
\end{align}
Each term describes the worldsheet of open strings that split and rejoin according to the target space cobordism.  However, in addition to the swapping of endpoints $(I,J)\rightarrow (I,L)$, $(K,L)\rightarrow (K,J)$, each stack of open strings experiences an interaction $(I,L)\rightarrow (I,\alpha(L))$,  $(K,J)\rightarrow (K,\sigma(J))$.  In the right figure, this whole process is compressed into a single interaction point located at the saddle point of the target manifold. 
We identify this as an $\Omega^{-1}$ point, because the worldsheet is branched over that point according to  $\sigma \alpha$ and is weighted with the appropriate factor of $\omega_{\sigma \alpha}$.

The target space for the density matrix $\rho= \psi \psi^{\dagger}$ is a sphere with four holes.  The corresponding effective partition function is 
\begin{align}
    Z=\text{tr} \rho = \sum_{n}  \sum_{\sigma,\alpha,\rho,\beta \in S_{n} } \sum_{\epsilon,\eta \in S_{n} } \frac{\omega_{\sigma \alpha}}{N^n} \frac{\omega_{\rho \beta}}{N^n}N^{K_{\epsilon}}N^{K_{\epsilon \beta \alpha }}N^{K_{\eta}} N^{K_{\eta \rho \sigma}}
\end{align}
Each term in this sum corresponds to open string worldsheets that end on $N$ entanglement branes at each hole.
The worldsheets encounter two $\Omega^{-1}$ points which is consistent with the Euler characteristic $\chi=-2$ of the target space.
Using the entanglement brane axiom, we can close up the entanglement branes into four $\Omega$ points.
The target space then becomes a sphere with $\chi=4-2$, consistent with the cancellation of $\Omega$ and $\Omega^{-1}$ points.  

The multi-interval case follows a similar pattern.  This enables us to give a worldsheet prescription for entanglement entropy in which we insert $2n$ stacks of $N$ entanglement branes at the entangling points, and $s = 2n - \chi$ interaction points in the bulk correponding to the $\Omega^{-1}$ points.

\section{Discussion}

By appealing to the framework of extended TQFT, we have shown how entanglement entropy in TQFT can be naturally understood in terms of an embedding of the closed Hilbert space into the open Hilbert space given by particular cobordisms.
We also showed that the language of cobordisms is naturally suited to describe modular flows of multiple disjoint regions and for computations of entanglement entropy and negativity.

It would be interesting to generalize this framework to understand entanglement in two-dimensional conformal field theories.
Here the entanglement brane axiom may have to be modified to account for the UV divergence that arises when closing up the entanglement boundary.  For example, this divergence appears in the leading term of the entanglement entropy of a single interval.

Nevertheless, for CFT states prepared from the Euclidean path integral, we expect that the entanglement entropy will probe the data of the open-closed CFT, which satisfies similar rules as the open-closed TQFT \cite{Lewellen:1991tb}.
Here, the relevant boundary conditions are conformal boundary conditions which satisfy the Cardy condition.
For a rational CFT, these correspond to finitely many Cardy states.
For the case of the Ising model CFT, \cite{Ohmori:2014eia} showed how the entanglement entropy depends on these Cardy states and how they are mapped to entanglement boundary conditions of the microscopic model.
In \cite{Wong:2018svs}, the multi-interval modular Hamiltonian was obtained by incorporating the cutting and gluing operations that are manifest in the cobordism of figure \eqref{saddle}.
These examples give some hints for how to incorporate entanglement calculations in the framework of extended CFT.

Another natural extension of the current work is to TQFT in higher dimensions.
We saw in section \ref{section:ebrane} that the entanglement brane axiom gives further constraints between the open and closed sectors and hence may simplify the classification of higher-dimensional TQFTs.
In particular, Chern-Simons theory in three dimensions provides a model where we expect a richer set of boundary conditions, since many edge theories are consistent with the same bulk \cite{Cano:2013ooa}.  
Here we will also encounter higher-codimension objects such as the interface of a physical boundary with the entanglement partition.  

We have reformulated the chiral Gross-Taylor string theory as an open-closed TQFT, without reference to 2D Yang-Mills.
This formulation shows that to cut the closed string into open strings in a way that is consistent with the Moore-Segal and entanglement brane axioms, we have to introduce $N= \frac{1}{g_{s}} $ Chan-Paton factors, corresponding to $N$ entanglement branes at each entangling surface.
It would be interesting to apply the same construction to the full non-chiral Gross Taylor string, where BV-BRST structure is expected to emerge due to constraints between the two chiral sectors \cite{Donnelly:2016jet}.

We have seen that the entanglement brane axiom implies that the closed sector knows about the density of states in the open sector.
Although the entanglement entropy of an interval in Yang-Mills theory includes a contribution from the edge modes, it can be calculated via the replica trick without reference to the open sector.
This is related to the fact that the Euclidean partition function of gravity can be used to find the Bekenstein-Hawking entropy without explicit reference to the underlying microstates being counted.
It has been argued that this is more than just an analogy, and that the Bekenstein-Hawking entropy could be understood as a contribution from edge modes, analogous to the term $\log \dim R$ appearing in the entanglement entropy of 2D Yang-Mills \cite{Donnelly:2016jet,Harlow:2016vwg,Lin:2017uzr}.
Further support for this relation has recently been found from holographic arguments \cite{Akers:2018fow,Dong:2018seb}.
They suggest a picture in which the gravitational Hilbert space of a region with boundary splits into sectors according to the area of the boundary, with states transforming in a representation of a local symmetry group of dimension $e^{A/4G}$.

A natural setting to understand this conjecture is Jackiw-Teitelboim gravity, where an independent counting of the gravitational microstates of a two-sided wormhole are provided by the dual Schwarzian quantum mechanics \cite{Harlow:2018tqv,Blommaert:2018oro}.
Jackiw-Teitelboim gravity admits a formulation as a two-dimensional BF theory closely related to the two-dimensional Yang-Mills theory considered here.
However it was pointed out in \cite{Lin:2018xkj} that naive application of the Yang-Mills results to the BF theory does not reproduce the results of the Schwarzian theory and hence does not lead to the correct formula for the Bekenstein-Hawking entropy.
Thus it remains an interesting open problem to understand whether Jackiw-Teitelboim gravity can be formulated as an open-closed TQFT satisfying the entanglement brane axiom.\footnote{Shortly after v1 of this paper appeared on the arxiv, Ref.~\cite{Blommaert:2018iqz} made significant progress on this question using a restricted version of the BF theory. }

\acknowledgments

GW would like to thank Janet Hung for discussions on category theory that was essential to this work.
WD thanks Bianca Dittrich and Laurent Freidel for discussions.
GW is supported by Fudan University and the Thousand Young Talents Program and thank Perimeter Institute and the Stanford Institute for Theoretical Physics for hospitality while this work was being finished. Research at Perimeter Institute is supported by the Government of Canada through the Department of Innovation, Science and Economic Development Canada and by the Province of Ontario through the Ministry of Research, Innovation and Science.

\appendix

\section{Derivations of open string cobordisms}
The main identity we need in computing cobordims for the CGT string is $\Omega \Omega^{-1} = \mathbf{1}$ in the $S_n$ algebra, which implies that
\begin{align} 
    \mathbf{1}&=\sum_{\mathbf{\alpha}, \mathbf{\beta} \in S_{n} } (N^{K_{\alpha}-n}  \omega_{\beta}) \,\, \mathbf{\alpha}\mathbf{ \beta} \nn
    &=   \sum_{\mathbf{\alpha}, \mathbf{\sigma} \in S_{n} } (N^{K_{\alpha}-n}  \omega_{\alpha^{-1} \sigma }) \,\,\mathbf{\sigma}
\end{align}
With a change of labeling this can be written as an identity
\begin{equation}\label{omegainv}
    \sum_{\mathbf{\alpha} \in S_{n} } (N^{K_{\alpha}-n}  \omega_{\alpha \sigma^{-1} }) = \delta (\sigma).
\end{equation}
For example, the co-unit $\epsilon$ is obtained from the gluing:
\begin{align}
\mathtikz{\pairA{0cm}{0cm}  \etaA{.5cm}{0cm} } &= \mathtikz{\epsilonA{0cm}{0cm}} \nn
\end{align}
Evaluating on a basis element, this gives
\begin{align}
    \epsilon(\ket{IJ}) &=\sum_{\tau, L} \sum_{\alpha, \sigma}  \frac{\omega_{\alpha} N^{K_{\tau} -n}}{n!} \delta_{I,\alpha \sigma \tau (L)} \delta_{J,\sigma(L)}\nn
    &=\sum_{\beta,\alpha} \omega_{\alpha} N^{K_{\beta}-n} \delta_{I,\alpha \beta(J)}\nn
    &= \sum_{\beta ,\rho } \omega_{\rho \beta^{-1} } N^{K_{\beta}-n} \delta_{I,\rho(J)}\nn
    &= \delta_{IJ} 
\end{align}

\paragraph{Cozipper}
The co-zipper is obtained from the following cobordism 
\begin{align}
    \mathtikz { \cozipper{0cm}{1cm}  } = \mathtikz { \pairA{0cm}{0cm} \zipper{-0.5cm}{1cm}  \idC{-1.5cm}{1cm} \copairC{-1cm}{1cm} \idA{.5cm}{1cm}}
\end{align}
which gives:
\begin{align} \label{cz}
    i^*(\ket{IJ}) &= \sum_n \frac{1}{n!} \sum_{\sigma,\alpha,\beta \in S_n} \frac{\omega_\alpha}{N^n} \delta(J, \alpha \beta \sigma \beta^{-1} I) \ket{\sigma} \\
    &= \sum_n \sum_{\alpha, \tau \in S_n} \frac{\omega_\alpha}{N^n} \delta(J, \alpha \tau I) \ket{\tau}
\end{align}

We can also verify the E brane axiom directly using the identity \eqref{omegainv}:
\begin{align}
\mathtikz{ \etaA{0cm}{0cm} \cozipper{0cm}{0cm} } &=\sum_{n,\alpha,\sigma \tau }   \frac{1}{n!}  \frac{\omega_{\alpha} N^{K_{\sigma}}}{N^{n}} \sum_{I} \delta_{I, \alpha \tau \sigma(I)}\ket{\tau}  \nn
  &=\sum_{n,\alpha,\sigma \tau }   \frac{1}{n!}  \omega_{\alpha} N^{K_{\sigma}} N^{K_{\alpha \tau \sigma}-n} \ket{\tau} \nn 
  &=\sum_{n,\sigma \tau }   \frac{1}{n!} N^{K_{\sigma}} \sum_{\beta} \omega_{\beta (\tau \sigma)^{-1}}  N^{K_{\beta}-n} \ket{\tau}\nn
  &=\sum_{n,\sigma \tau }   \frac{1}{n!} N^{K_{\sigma}}  \ket{\sigma}=\mathtikz{\etaC{0cm}{0cm}}.
\end{align}

\bibliographystyle{utphys}
\bibliography{open-closed}

\end{document}